\documentclass[aps,prd,amsmath,amssymb,nofootinbib,eqsecnum, preprintnumbers,twocolumn,longbibliography]{revtex4-1}

\usepackage{hyperref}
\usepackage{graphicx,xcolor}
\usepackage{xcolor}

\usepackage[scr=boondoxo]{mathalfa}

\def\pa{{\partial}}
\newcommand{\be}{\begin{equation}}
\newcommand{\ee}{\end{equation}}
\newcommand{\ba}{\begin{eqnarray}}
\newcommand{\ea}{\end{eqnarray}}
\newcommand{\nn}{\nonumber}

\newcommand{\lagr}{\mathscr{l}}
\newcommand{\ham}{\mathscr{h}}

\newcommand{\pix}[1]{{\mathsf{#1}}}

\newcommand{\obsix}[1]{{\mathchoice{\displaystyle #1}{\textstyle #1}{\textstyle #1}{\scriptstyle #1}}}

\begin{document}
\title{{On Integrability of the Geodesic Deviation Equation}}

\author{{Marco Cariglia}}
\email{marco.cariglia@ufop.edu.br}
\affiliation{Departamento de F\'isica, Universidade Federal de Ouro Preto, 35400-000 Ouro Preto MG, Brazil}

\author{{Tsuyoshi Houri}}
\email{t.houri@maizuru-ct.ac.jp}
\affiliation{National Institute of Technology, Maizuru College, 234 Shiraya, Maizuru, Kyoto 625-8511, Japan}

\author{{Pavel Krtou\v{s}}}
\email{Pavel.Krtous@utf.mff.cuni.cz}
\affiliation{Faculty of Mathematics and Physics, Charles University,\\
V~Hole\v{s}ovi\v{c}k\'ach 2, Prague, Czech Republic}

\author{{David Kubiz\v{n}\'ak}}
\email{dkubiznak@perimeterinstitute.ca}
\affiliation{Perimeter Institute, 31 Caroline St. N. Waterloo
Ontario, N2L 2Y5, Canada}

\date{May 19, 2018}  

\begin{abstract}
The Jacobi equation for geodesic deviation describes finite size effects due to the gravitational tidal forces. In this paper we show how one can integrate the Jacobi equation in any spacetime admitting completely integrable geodesics. Namely, by linearizing the geodesic equation and its conserved charges, we arrive at the invariant Wronskians for the Jacobi system that are linear in the `deviation momenta' and thus yield a system of first-order differential equations that can be integrated. The procedure is illustrated on an example of a rotating black hole spacetime described by the Kerr geometry and its higher-dimensional generalizations. A number of related topics, including the phase space formulation of the theory and the derivation of the covariant Hamiltonian for the Jacobi system are also discussed.

\vspace{0.2\baselineskip} \noindent\textbf{Keywords}:
Geodesics, Jacobi field, Integrability, Killing tensors, Black holes.
\end{abstract}

\maketitle

\section{Introduction}

The Jacobi equation of geodesic deviation, e.g. \cite{Wald:book1984}, has an important place in General Relativity. The motion of a small enough particle with no internal structure such as charge or spin is described by a geometrical object, a geodesic, so that by the principle of equivalence an observer freely falling with the particle would not be able to feel the effects of gravity in a small neighbourhood of spacetime. Gravitational effects will become apparent considering objects of a finite size, whose evolution can be thought of as that of a bundle of nearby geodesics: at first order in the separation parameter between the geodesics these effects are described by the Jacobi equation. This happens for example when studying the effects of a passing gravitational wave, as in the gravitational-wave memory effect \cite{Favata:2010zu,Duval:2017els,Zhang:2017rno,Zhang:2017geq,Zhang:2018srn}. The Jacobi equation can be generalized to the case of particles with electric charge \cite{Balakin:2000wt}, particles with spin \cite{vanHolten:2001ea}, or to the non-linear case where the dependence on relative velocities is not linearized \cite{chicone:2002}.

The geodesic deviation has been used to
give a geometrical and physical interpretation of spacetimes in ordinary four dimensions and higher \cite{podolsky:2012,Podolsky:2013ola,Svarc:2011zz}, while using first and higher order it has been used to construct approximations to generic geodesics starting from simple ones  \cite{Kerner:2001cw,Colistete:2002ka}, that can be used to model extreme mass-ratio systems \cite{Koekoek:2011mm}. In another approach, it has been used to study geodesic (in)stability for dynamical systems, and Lyapunov exponents \cite{Casetti:2000gd,Pettini:book2007}.

In this work we analyze the Jacobi equations from the point of view of Hamiltonian dynamics and symmetries of the dynamics. The Jacobi dynamical system is non-trivial, even though it is linear, because of its time dependence and the complexity of its differential equations. Time dependent Hamiltonian systems have been recently discussed in \cite{cariglia:2016} in the light of the Eisenhart lift technique \cite{duval:1985,duval:1991,Cariglia:2014ysa,Cariglia:2018mos}. On the other hand, due to their complexity, few explicit solutions of the equations are given in the literature \cite{dryuma:1997,Balakin:2000wt,Kerner:2001cw,philipp:2015}.

From this point of view it is interesting that hidden symmetries of the related geodesic equations generate non-trivial solutions of the Jacobi equation \cite{Caviglia:1982a,Caviglia:1983,dolan:1984}, displaying the relationship between the Jacobi equations and the concept of hidden symmetries.

In fact the equations of geodesic deviation are naturally linked to those of geodesic motion. In \cite{bazanski:1989} a method is presented that generalizes the Hamilton--Jacobi equation, allowing to obtain at the same time solutions of the geodesic and the geodesic deviation equations. A Lagrangian formulation of the geodesic deviation equations, including an electromagnetic field and spin, and a treatment of higher order deviation equations can be found in \cite{Balakin:2000wt,Kerner:2001cw,kerner:2003}, and is obtained by an expansion of that of geodesic motion. Given this natural connection, it is reasonable to expect that symmetries of dynamics of the geodesic motion, when present, descend to symmetries of the Jacobi equation. This is the focus of the present work, where we show that integrals of geodesic motion give rise, through linearization, to integrals of the Jacobi equation that are expressed in the form of invariant Wronskians. Of particular interest are the geodesic integrals built from Killing vectors and Killing tensors, which are associated to hidden symmetries of the geodesic dynamics and that generate integrals for the Jacobi equation. The latter integrals  inherit an algebraic structure via Poisson brackets that is isomorphic to the structure of the geodesic integrals. In particular, if in $n$ dimensions there are $n$ functionally independent, mutually Poisson commuting integrals of the geodesic motion, then these induce via linearization a set of $n$ independent, mutually commuting integrals for the Jacobi equation, thus showing that integrability of the geodesic equations implies integrability of the Jacobi equations.

The structure of the work is as follows.  We begin in Sec.~\ref{sec:geodesic} by reviewing the concept of geodesic motion and setting some of the notation. In Sec.~\ref{sec:Jacobi} we present the Jacobi equation and discuss its Lagrangian and Hamiltonian formulation. We first discuss a coordinate approach where the Hamiltonian is not a scalar with respect to changes of coordinates in the base manifold: in this case there is a coordinate Hamiltonian for each coordinate chart, different such Hamiltonians  being related by canonical transformations in overlapping regions, and global motion is obtained sewing solutions from different charts. We then introduce a covariant Hamiltonian approach where the Hamiltonian is a globally defined scalar. In Sec.~\ref{sc:symdyn} we discuss how geodesic integrals of motion descend to integrals of the Jacobi equation. We present the notion of invariant Wronskians, followed by the conserved quantities generated by Killing tensors and then discuss integrability. In Sec.~\ref{sec:black_holes} we apply our results to rotating black holes in four dimensions, and mention how these generalize to higher dimensional Kerr--NUT--(A)dS black holes. The Jacobi equation is integrable in these geometries and from our results it is possible to build a complete set of mutually commuting conserved charges. Sec.~\ref{sec:conclusions} presents concluding remarks and possible future lines of research. Appendix~\ref{apx:PhsSpc} presents the Jacobi equation, the linearization procedure and the Wronskians from the point of view of a general phase space and a general set of equations of motion, before the introduction of a cotangent bundle or a specific Hamiltonian, and Appendix~\ref{apx:CovLin} details the construction of a covariant Lagrangian and Hamiltonian.

\section{Geodesic motion}
\label{sec:geodesic}

Let $\mathcal{M}$ be a spacetime manifold of dimension $n$, equipped with metric $g_{ab}$, and Riemann tensor\footnote{%
{The convention we use in this work for the commutation of covariant derivatives, when acting for example on a vector $V^a$, is
\[
\left[ \nabla_a , \nabla_b \right] V^c =  R_{ab}{}^c{}_d V^d \, .
\]}
$\frac{D}{D\tau}$ represents the covariant derivative and\,  $\frac{d}{d\tau}=\dot{}\ $\, a coordinate derivative along a curve.
}
$R_{abcd}$. Having in mind applications to relativity we take the metric to be of almost plus type, though with minor adjustments everything would readily  generalize to a generic metric.

There are several approaches to the Lagrangian and Hamiltonian descriptions of the relativistic particle. We follow the approach used by Carter \cite{Carter:1968cmp} and others , where one identifies the spacetime $\mathcal{M}$ with a configuration space of the system. The particle is described by its trajectory $x^a(\lambda)$ parametrized by an external time $\lambda$. The phase space is then described by the cotangent space $\mathbf{T}^*\mathcal{M}$. The Lagrangian for geodesic motion is given by
\begin{equation}\label{eq:geodLagr}
    L = \frac12g_{ab}(x)\frac{d x^a}{d\lambda}\frac{d x^b}{d\lambda}\;.
\end{equation}
The momentum and the Hamiltonian read
\begin{equation}\label{eq:geodmom}
    p_a = g_{ab} \frac{d x^b}{d\lambda}\;,
\end{equation}
\begin{equation}\label{eq:geodHam}
    H = \frac12 g^{ab}(x) p_a p_b\;,
\end{equation}
and the Hamilton equations reduce to \eqref{eq:geodmom} and
\begin{equation}\label{eq:geodmomconst}
    \frac{D p_a}{D\lambda}=0\;.
\end{equation}
This gives, of course, the geodesic equation.

This formalism describes a free particle of an arbitrary mass. (In this paper we concentrate on the case of massive particles for which $dx^a/d\lambda$ is not a null vector.) The mass is fixed by the value of the Hamiltonian, i.e., by the normalization of the momentum
\begin{equation}\label{eq:massconstr}
    H = -\frac12 m^2\;.
\end{equation}
Since the Hamiltonian does not depend on the external time explicitly, it is a conserved quantity and the mass is for a given trajectory fixed.

We can always rescale the time variable and introduce the proper time $\tau$
\begin{equation}\label{eq:propertime}
    \tau = m\lambda\;,
\end{equation}
with the normalization of the velocity $u^a$
\begin{equation}\label{eq:velnorm}
    u^a = \frac{dx^a}{d\tau}=\frac{p^a}{m}\;,\quad u^a u^b g_{ab} = -1\;.
\end{equation}
Of course, free particles of different (non-zero) masses follow the same geometric trajectories---they differ only in time parametrization. We can thus ignore a particular value of the mass and describe the geodesic in proper time parametrization.

In what follows we consider a situation where the geodesic motion admits a nontrivial number of integrals of motion.
In particular, the {conserved quantities} that are homogeneous in particle's momentum,
\be \label{eq:Killing_charge}
K(x,p) = \frac{1}{r!}\, K^{a_1 \dots a_{r}}(x)\, p_{a_1} \dots p_{a_{r}} \,,
\ee
are in one to one correspondence with Killing tensors \cite{Walker:1970un}. A  Killing tensor of rank $r$ is a symmetric tensor $K^{a_1 \dots a_r}=K^{(a_1 \dots a_r)}$, such that
\be\label{eq:KillingTens}
\nabla^{(a_0} K^{a_1 \dots a_{r})} = 0 \, .
\ee
For $r=1$ it reduces to the Killing vector.

It is the purpose of the present paper to linearize these conserved quantities and show that they give rise to simple integrals of motion for the Jacobi geodesic deviation equation described in the next section. In particular, it implies that when the original geodesic motion is completely integrable, so will be the Jacobi equation.

\section{The Jacobi system\label{sec:Jacobi}}
\label{sc:JacobiEq}

\subsection{Geodesic deviation equation}

Given a geodesic  $\bar{x}^a(\tau)$, we want to study its nearby trajectories. To that purpose we consider a one-parameter family of curves $x^a(\sigma, \tau)$, with $\sigma$ being the parameter labeling different curves and $\tau$ the time parameter along each curve. We call the geodesic $\bar{x}^a(\tau)=x^a(0,\tau)$ a central geodesic. We assume $\tau$ to be the proper time along the central geodesic, however, it does not have to be the proper time for trajectories with nonvanishing $\sigma$. Nevertheless, we still denote the velocity with respect to $\tau$ as $u^a$,
\begin{equation}\label{eq:velfam}
    u^a(\sigma,\tau) = \frac{d x^a}{d\tau}(\sigma,\tau)\;.
\end{equation}
We call by $n^a$ a vector that links the nearby curves,
\be
n^a (\sigma,\tau) = \frac{dx^a}{d\sigma}(\sigma,\tau)\;.
\ee
For $\sigma=0$, $\bar{u}^a(\tau)\equiv u^a(0,\tau)$ reduces to the normalized velocity of the central geodesic and $n^a(\tau) \equiv n^a(0,\tau)$ describes the deviation from the central geodesic. One can think of this vector as a linear approximation for the trajectories close to the central geodesic.

A special degenerate case occurs when the whole family ${x^a(\tau,\sigma)}$ lies entirely on the central geodesic. Clearly, $n^a$ is then proportional to $u^a$.

Specifically, we are interested in the trajectories in the vicinity of the central geodesic which are {geodesics as well}, i.e., curves that (for all values of the parameter $\sigma$) satisfy
\be \label{eq:geodesics}
\frac{D^2 x^a}{D \tau^2} = \frac{d^2 x^a}{d\tau^2} + \Gamma^a_{bc} \frac{d x^b}{d\tau} \frac{d x^c}{d\tau} = 0 \, .
\ee
In this case {it is well known that} the deviation $n^a(\tau)$ connecting nearby geodesics must satisfy the geodesic deviation Jacobi equation,
\be \label{eq:Jacobi}
\frac{D^2 n^a}{D \tau^2} + \bar{R}^a{}{}_{cbd}\, \bar{u}^c \bar{u}^d n^b = 0 \, .
\ee
Here, the bar above the Riemann tensor (and, similarly, above other quantities) indicates that it is evaluated at the central geodesic, $\bar{R}_{abcd} = R_{abcd}(\bar{x})$.

In order to derive this equation, the key observation is to realize that since $u^a$ and $n^a$ are essentially coordinate vectors, $u\equiv\pa_\tau$, $n\equiv\pa_\sigma$, on the 2-surface $x^a(\sigma,\tau)$, they Lie-commute, $[u,n]=0$, which when expressed in terms of the metric covariant derivative yields
\be
\frac{D n^a}{D\tau} = \frac{D u^a}{D\sigma}\,,
\ee
see e.g. \cite{Wald:book1984} for more details.

\subsection{Lagrangian for the Jacobi equation}

The Jacobi equation admits a Lagrangian formulation, with the Lagrangian
${\lagr}$ given by \cite{Balakin:2000wt}
\be \label{eq:Lagrangian_vanHolten}
\lagr = \frac{m}{2} \bar{g}_{ab} \frac{D n^a}{D \tau} \frac{D n^b}{D \tau}
   - \frac{m}{2} \bar{R}_{abcd}\, \bar{u}^a \bar{u}^c n^b n^d \,.
\ee
This is the Lagrangian for independent `deviation' variable $n^a$ which represents a general curve close to the central geodesic. As we will discuss below, it can be understood as a function $\lagr(n^a,\dot n^a)$ of coordinate velocity $\dot n^a=\frac{dn^a}{d\tau}$, or it can be treated `covariantly', as a function $\lagr(n^a, \frac{Dn^a}{d\tau})$ of covariant velocity~$\frac{Dn^a}{d\tau}$.
In either case the corresponding Euler--Lagrange equations pick up a nearby geodesic specified by the Jacobi equation.

The Lagrangian \eqref{eq:Lagrangian_vanHolten} can be derived by the linearization process, starting from the Lagrangian of the geodesic motion.\footnote{
The appearance of mass $m$ in the Lagrangian \eqref{eq:Lagrangian_vanHolten} comes from replacing the external time $\lambda$ with the proper time $\tau$ along the central geodesic. The linearization starts from the full Lagrangian \eqref{eq:geodLagr} multiplied by an additional factor $m^{-1}$ coming from the integration element in the action $S=\int L\, d\lambda=\int L\, m^{-1}d\tau$.
}
We refer to the  Appendix~\ref{apx:CovLin} for the derivation and further technical details.

The Lagrangian \eqref{eq:Lagrangian_vanHolten} is obviously time dependent---it depends on the time-dependent position $\bar{x}(\tau)$ and velocity $\bar{u}(\tau)$ of the central geodesic. Therefore, there is one such Lagrangian for each central geodesic. While the dependence on the velocity $\bar{u}(\tau)$ is explicit, the dependence on $\bar{x}(\tau)$ enters through the spacetime dependence of the metric, Christoffel symbols, and the Riemann tensor, and to stress this fact we write these objects with bar. All these should be considered as given functions of the time variable $\tau$ and provide apriori data for the Jacobi equation. Of course, for a concrete spacetime, it may be difficult to obtain the expression for
$\bar{x}(\tau)$ and $\bar{u}(\tau)$ in an explicit and closed form. We return to the question of time-dependency of the Lagrangian again below, when we discuss it from a covariant perspective.

\subsection*{Transverse and tangent splitting}

Before we proceed to the Hamiltonian formulation, let us first discuss decoupling of the transverse and parallel degrees of freedom in the deviation variable. The Jacobi equation \eqref{eq:Jacobi} always admits the following two solutions:
\be \label{eq:two_trivial_solutions}
\begin{aligned}
  {}^1n^a(\tau) &= \bar{u}^a \, ,\\
  {}^2n^a(\tau) &= \tau \bar{u}^a \, ,
\end{aligned}
\ee
which arise from the degenerate case when the entire family $x^a(\sigma,\tau)$ lies on the central geodesic $x^a(\tau)$:
\be\label{eq:two_trivial_families}
\begin{aligned}
 {}^1x^a (\sigma;\tau) &= x^a (\tau + \sigma)\,, \\
 {}^2x^a (\sigma;\tau) &= x^a (e^\sigma\tau ) \, .
\end{aligned}
\ee
These solutions of the Jacobi equation correspond to the known freedom of redefining the affine parameter. Since Eq.~\eqref{eq:Jacobi} is a linear equation of the second order, in an $n$-dimensional spacetime the space of solutions is $2n$-dimensional. Removing the trivial solutions \eqref{eq:two_trivial_families} amounts to reducing the problem to the subspace of deviation vectors that are transverse to the geodesics. Namely, writing
\be
  n^a = n_\parallel^a + n_\perp^a\,, \quad
  n_\parallel^a = \nu(\tau) \bar{u}^a\,, \quad
  g_{ab}\, n_\perp^a \bar{u}^b =0\,,
\ee
we obtain two separate equations
\begin{gather}
\frac{D^2 n_\parallel^a}{D \tau^2} = 0 \,,\label{eq:Jacobi_parallel}\\
\frac{D^2 n_\perp^a}{D \tau^2} + \bar{R}^a{}_{cbd} \bar{u}^c \bar{u}^d n_\perp^b = 0 \,. \label{eq:Jacobi_perp}
\end{gather}
The solutions of \eqref{eq:Jacobi_parallel} are precisely those in Eq.~\eqref{eq:two_trivial_solutions}.

Correspondingly, for timelike $\bar{u}^a$, the Lagrangian separates as $\lagr = \lagr_\parallel + \lagr_\perp$, with
\begin{align}
\lagr_\parallel &= \frac{m}{2} \bar{g}_{ab}  \frac{D n_\parallel^a}{D \tau} \frac{D n_\parallel^b}{D \tau} \, ,  \\
\lagr_\perp &= \frac{m}{2} \bar{g}_{ab} \frac{D n_\perp^a}{D \tau} \frac{D n_\perp^b}{D \tau}
  - \frac{m}{2} \bar{R}_{abcd} \bar{u}^a \bar{u}^c n_\perp^b n_\perp^d \,, \label{eq:Lagrangian_perp}
\end{align}
as follows from the fact that
\be
\bar{g}_{ab} \frac{D n_\parallel^a}{D \tau} \frac{D n_\perp^b}{D \tau}
  = \bar{g}_{ab} \, \dot{\nu} \bar{u}^a \frac{D n_\perp^b}{D \tau} =
\dot{\nu} \frac{d}{d \tau} \left( \bar{g}_{ab} \bar{u}^a n_\perp^b \right) = 0 \, .
\ee
When the geodesics are timelike, the metric can be decomposed
\be
\bar{g}_{ab} = - \bar{u}_a \bar{u}_b + \bar{g}_{\perp ab} \, ,
\ee
where $\bar{g}_{\perp ab}$ is the projector to the transverse space, which is positive definite, and \eqref{eq:Lagrangian_perp} can be written as
\be
\lagr_\perp = \frac{m}{2} \bar{g}_{\perp ab} \frac{D n_\perp^a}{D \tau} \frac{D n_\perp^b}{D \tau}
  - \frac{m}{2} \bar{R}_{abcd} \bar{u}^a \bar{u}^c n_\perp^b n_\perp^d \, . \label{eq:Lagrangian_perp_2}
\ee

\subsection*{Hamiltonian formalism: coordinate approach}

In what follows we shall study the Jacobi equation from the point of the Hamiltonian dynamics using both, coordinate (this subsection) and covariant (next subsection) approaches.

In the coordinate approach, the Lagrangian \eqref{eq:Lagrangian_vanHolten} is understood as a function of the `positions' $n^a$ and the `coordinate velocities'~$\dot{n}^a$,
\be \label{eq:Lagrangian_coor}
\lagr {=} \frac{m}{2} \bar{g}_{ab} (\dot{n}^a{+}u^k\bar{\Gamma}^a_{kc} n^c) (\dot{n}^b{+}u^l\bar{\Gamma}^b_{ld} n^d) - \frac{m}{2} \bar{R}_{kalb} \bar{u}^k \bar{u}^l n^a n^b  .
\ee
To write down the corresponding Hamiltonian formulation, we define the momentum canonically conjugated to $n^a$
\be
  \pi_a = \frac{\pa \lagr}{\pa \dot{n}^a}
  = m \bar{g}_{ab} \left( \dot{n}^b + \bar{u}^k\bar{\Gamma}^b_{kc}  n^c \right)
  = m \bar{g}_{ab} \frac{D n^b}{D \tau} \, ,
\ee
and introduce what we call the coordinate Hamiltonian, ${\ham_{\mathrm{c}} = \pi_a \dot{n}^a - \lagr}$,
\be \label{eq:Hcoor}
\ham_{\mathrm{c}} = \frac{1}{2m} \bar{g}^{ab} \pi_a \pi_b - \bar{u}^k\bar{\Gamma}^a_{kb} \, \pi_a n^b
  + \frac{m}{2} \bar{R}_{abcd} \,  \bar{u}^a \bar{u}^c n^b n^d \, .
\ee
An explicit calculation shows that the equations of motion obtain from $\ham_{\mathrm{c}}$ imply the Jacobi equation \eqref{eq:Jacobi}.

It is obvious that the Hamiltonian \eqref{eq:Hcoor} is not covariant: it depends in a non-tensorial way on a particular choice of coordinates through the Christoffel symbols $\Gamma^a_{bc}$, reflecting the fact that we used a non-covariant form of the velocity $\dot{n}^a$. Such a velocity does not transform as a vector under a coordinate transformation and therefore the related Hamiltonian is also non-covariant. The Hamiltonian \eqref{eq:Hcoor} generates the coordinate time-evolution of vector quantities.

As we will see in the next subsection, it is possible to proceed in a more covariant way, starting with the covariant velocity $\frac{D n^a}{D\tau}$, and arrive at a simpler covariant Hamiltonian \eqref{eq:covHam} below. This, however, requires an additional care about technical details and before we explore such approach, let us first discuss the behavior of the coordinate Hamiltonian under a change of coordinates.

Evolving the parameter $\tau$, there can come a moment when the geodesic leaves the given coordinate chart $\{ x^a \}$. Then it is necessary to use a different set of coordinates
\be
x^{\prime a} = x^{\prime a} (x^b) \, , \qquad J^a{}_b = \frac{\pa x^{\prime a}}{\pa x^b} \, . \label{eq:change_coord}
\ee
From the point of view of Hamiltonian dynamics, such a change of coordinates induces a time-dependent canonical transformation of $(n^a, \pi_a)$ variables, and the Hamiltonian in general changes. The non-invariance property is expressed by the non-tensorial term in $\ham_{\mathrm{c}}$. To be explicit, the change of coordinate chart~\eqref{eq:change_coord} induces the following transformation:
\be\label{eq:canonical}
\begin{aligned}
n^{\prime a} &= J^a{}_b\, n^b \, ,  \\
\pi^\prime_a &= \pi_b\, J^{-1}{}^b {}_a \, .
\end{aligned}
\ee
Such a transformation is canonical and amounts to the following generating function:
\be
G(n,\pi^\prime) =  \pi^\prime_a \, J^a{}_b (\tau) \, n^b \, .
\ee
By this we mean that Eqs.~\eqref{eq:canonical} are obtained via solving
\be
\begin{aligned}
n^{\prime a} &= \frac{\partial G}{\partial \pi^\prime_a}(n,\pi^\prime) \, , \\
\pi_a &= \frac{\partial G}{\partial n^a}(n,\pi^\prime)
\end{aligned}
\ee
for $n^\prime$ and $\pi^\prime$. The standard theory then gives that the Hamiltonian transforms to
\be
\begin{split}
&\ham_{\mathrm{c}}^\prime (n^\prime, \pi^\prime) = \ham_{\mathrm{c}}(n,\pi) + \frac{\partial G}{\partial \tau}(n,\pi^\prime)\\
&\quad  =\frac{1}{2m} \bar{g}^{ab} \pi^\prime_a \pi^\prime_b
- \bar{u}^{\prime k} \bar{\Gamma}^{\prime a}_{kb}\, \pi^\prime_a  n^{\prime b}
+ \frac{m}{2} \bar{R}^\prime_{abcd} \bar{u}^{\prime a} \bar{u}^{\prime c} n^{\prime b} n^{\prime d} \, ,
\end{split}\raisetag{8ex}
\ee
where the term $\frac{\partial G}{\partial \tau}$ provides the non-tensorial term in the transformation rule of the connection. The Hamiltonian is thus locally invariant in form, although it is not build from pure tensorial expressions. However, there is no such thing as a `global coordinate Hamiltonian'. For each coordinate chart there is a different Hamiltonian which governs time evolution in that chart. These Hamiltonians differ by $\frac{\partial G}{\partial \tau}$ terms, and so they cannot be understood as just different coordinate expressions for one Hamiltonian function. The global motion has to be sewed from solutions in different charts. Nevertheless, the global covariant Hamiltonian can be defined in the covariant approach as we will show next.

\subsection*{Hamiltonian formalism: covariant approach}

The linearized configuration space of vectors $n^a$ is time dependent: it is a tangent space at $\bar{x}(\tau)$. In the discussion of the coordinate Hamiltonian above, we have implicitly identified such tangent spaces by choosing particular coordinates. Namely, we have naturally identified vectors with the same components with respect to the coordinate frame. More precisely, we regarded such vectors as `not changing'. We have used the coordinate-time derivative $\dot{n}^a= \frac{d n^a}{d\tau}(\tau)$ to define the velocity and employed it in the construction of the Hamiltonian. Since such an identification of vectors is non-covariant, we obtained a non-covariant Hamiltonian which was dependent explicitly on the Christoffel symbols.

However, we can identify the linearized configuration spaces at different times in a more covariant way: by a parallel transport along the central geodesic. For that, we use the covariant time derivative $\frac{D n^a}{D\tau}$ to define the velocity.\footnote{Such a procedure could be also described in terms of an orthogonal frame parallel-transported along the central geodesic. We then say that vectors at different times are the same (not changing) if they have the same components with respect to such a frame.}

Thus, in the covariant approach we must interpret the Lagrangian $\lagr$ as a function of a \emph{linearized position $n^a$} and of a \emph{covariant velocity $v^a = \frac{D n^a}{D\tau}$},
\begin{equation}\label{eq:Lnv}
   \lagr(n,v) = \frac12m\, \bar{g}_{ab} v^a v^b
                  - \frac12 m\, \bar{u}^c \bar{u}^d \,\bar{R}_{cadb}\, n^a n^b \;.
\end{equation}

A covariant version of canonically conjugate momentum $\pi_a$ then reads
\begin{equation}\label{eq:covmom}
    \pi_a = \frac{\partial \lagr}{\partial v^a} = m\, \bar{g}_{ab} v^b\,,
\end{equation}
and the covariant Hamiltonian, $\ham = \pi_a v^a - \lagr$, is
\begin{equation}\label{eq:covHam}
   \ham(n,\pi) 
      = \frac1{2m}\, \bar{g}^{ab} \pi_a \pi_b
        + \frac12 m\, \bar{u}^c \bar{u}^d \,\bar{R}_{cadb}\, n^a n^b \;.
\end{equation}

The Hamilton equations have to be written again using the covariant time derivative
\begin{equation}\label{eq:covHE}
\begin{aligned}
    \frac{D n^a}{D\tau} &= \frac{\partial \ham}{\partial \pi_a} = \frac1{m}\, \bar{g}^{ab}\pi_b\;,\\
    \frac{D\pi_a}{D\tau} &= - \frac{\partial \ham}{\partial n^a} =
      -m\, \bar{u}^c \bar{u}^d \,\bar{R}_{cadb}\, n^b \;.
\end{aligned}
\end{equation}
Combining both equations together yields the Jacobi equation,
\begin{equation}\label{eq:JacobifromHE}
    \frac{D^2 n^a}{D\tau^2} + \bar{R}^{a}{}_{cbd}\,\bar{u}^c \bar{u}^d\, n^b =0\;.
\end{equation}
The evolution of a general phase space observable $A$ expressed in new variables $n$, $\pi$ is given by
\begin{equation}\label{eq:Aevol}
    \frac{d}{d\tau}{A} = \{A,\ham\} + \frac{\pa A}{\pa \tau}\;,
\end{equation}
where the Poisson bracket assumes the standard form
\begin{equation}\label{eq:PNnpi}
    \{A,B\} = \frac{\pa A}{\pa n^c}\frac{\pa B}{\pa \pi_c} - \frac{\pa A}{\pa \pi_c}\frac{\pa B}{\pa n^c}\;.
\end{equation}

Let us stress that even the observables, that were independent of the time parameter
in the original variables $(x,\,p)$, typically become explicitly time dependent and the second term in \eqref{eq:Aevol} is non-trivial. The reason is that a `simple' function of $x$ and $p$ is re-expressed in terms of the central trajectory $\bar{x}$, $\bar{p}$ and linearized variables $n$ and $\pi$. The explicit time dependency then enters through the time dependent central trajectory.

In particular, this is true for the covariant Hamiltonian itself, and we have
\begin{equation}\label{eq:hamder}
    \frac{d}{d\tau}\ham = \frac{\pa\ham}{\pa\tau}\;,
\end{equation}
that is, the new Hamiltonian $\ham$ for the linearized system is not conserved. However, for the Hamiltonian, the time dependency is not caused only by the introduction of the linearized variables with respect to the central trajectory, but it is also related
to a time-dependent canonical transformation that relates the original Hamiltonian $H$ and its (linearized) version  $\ham$, see appendix~\ref{apx:CovLin} for more details.

\section{Integrals of motion for the Jacobi system}
\label{sc:symdyn}

In this section we will describe how the integrals of motion for geodesics give rise to the particularly simple integrals of motion for the Jacobi system.
We start with a discussion applicable to any linearized dynamical system and only later specify to the case of the geodesic motion.

\subsection{Wronskian as a linear integral of motion}

Consider a linearized dynamical system, with trajectories near the central trajectory $\bar{x}$ described by a linearized trajectory $n^a(\tau)$. A general feature of linearized systems is that their dynamics is governed by a quadratic Lagrangian $\lagr(n,v)$, and, in the Hamiltonian picture, by a quadratic Hamiltonian $\ham(n,\pi)$, see \eqref{eq:Lnv} and \eqref{eq:covHam} for the specific example of linearized geodesic motion.

For two linearized trajectories $n_1^a(\tau)$ and $n_2^a(\tau)$, we define the Wronskian as
\begin{equation}\label{eq:Wronskiandef}
    W[n_1|n_2] = n_1^a \frac{\pa \lagr}{\pa v^a}\Bigl(n_2,\frac{D n_2}{D\tau}\Bigr)
     - \frac{\pa \lagr}{\pa v^a}\Bigl(n_1,\frac{D n_1}{D\tau}\Bigr) n_2^a\;.
\end{equation}
In the phase-space variables the trajectory is characterized by position $n^a$ and momentum $\pi_a$ and the Wronskian can be written as
\begin{equation}\label{eq:WronskianJacnpi}
    W[n_1,\pi_1|n_2,\pi_2] = n_1^a \pi_{2a} - \pi_{1a} n_2^a\;.
\end{equation}

It is well known (see, e.g., \cite{Caviglia:1982a,Caviglia:1983} for the case of the Jacobi system) that for any two solutions $n_1^a(\tau)$ and $n_2^a(\tau)$ of the equation of motion the Wronskian is conserved in time $\tau$. To show this, let us use the Hamiltonian picture. By employing the general quadratic Hamiltonian,
\begin{equation}\label{eq:hamquadr}
    \ham(n,\pi) = \frac12 \pi_a \bar{K}^{ab}\pi_b + \pi_a \bar{A}^a{}_b n^b + \frac12 n^a \bar{U}_{ab} n^b\;,
\end{equation}
we have the following Hamiltonian equations:
\begin{equation}\label{eq:covHEquadr}
\begin{aligned}
    \frac{D n^a}{D\tau} &= \frac{\partial \ham}{\partial \pi_a}
       = \bar{K}^{ab}\,\pi_b + \bar{A}^a{}_b\, n^b\;,\\
    \frac{D\pi_a}{D\tau} &= - \frac{\partial \ham}{\partial n^a} =
      -\, \bar{U}_{ab}\, n^b - \pi_b\, \bar{A}^b{}_a \;.
\end{aligned}
\end{equation}
Taking the time derivative of \eqref{eq:WronskianJacnpi} and substituting \eqref{eq:covHEquadr} for $\frac{D n_1}{D\tau}$, $\frac{D\pi_1}{D\tau}$ and $\frac{D n_2}{D\tau}$, $\frac{D\pi_2}{D\tau}$, we find
\be
\frac{d}{d\tau}W[n_1,\pi_1|n_2,\pi_2] =0\,.
\ee

This means that any fixed solution $\tilde{n}(\tau)$ generates a quantity
\begin{equation}\label{eq:consWr}
    W_{\tilde{n}}(n) = W[\tilde{n}|n]\,,
\end{equation}
which is conserved along any solution $n(\tau)$. In the phase-space language, any solution $\tilde{n}(\tau)$, $\tilde{\pi}(\tau)$ defines a conserved quantity
\begin{equation}\label{eq:consWrnpi}
    W_{\tilde{n},\tilde{\pi}}(n,\pi) = W[\tilde{n},\tilde{\pi}|n,\pi]
      = \tilde{n}^a \pi_{a} - \tilde{\pi}_{a} n^a\;.
\end{equation}
Clearly, such a conserved quantity is linear in $n$ and $\pi$.

This observation can be reversed: the most general linear conserved quantity of a linearized system is given by $W_{\tilde{n},\tilde{\pi}}$, where $\tilde{n}(\tau)$, $\tilde{\pi}(\tau)$ are explicit solutions of the Hamilton equations.
Indeed, let us consider a general linear observable ${C=\tilde{n}^a \pi_{a} - \tilde{\pi}_{a} n^a}$ with yet unspecified coefficients $\tilde{\pi}_{a}(\tau)$ and $\tilde{n}^{a}(\tau)$. Its conservation means
\begin{equation}\label{eq:Ccons1}
\begin{split}
    0&=\frac{d}{d\tau}C(n,\pi) = \{C,\ham\} + \frac{\pa C}{\pa\tau}\\
     &=-\tilde{\pi}_a\frac{\pa\ham}{\pa\pi_a}(n,\pi) - \tilde{n}^a \frac{\pa\ham}{\pa n^a}(n,\pi)
       + \frac{D\tilde{n}^a}{D\tau} \pi_a - \frac{D\tilde{\pi}_a}{D\tau} n^a \;.
\end{split}\raisetag{8ex}
\end{equation}
Substituting \eqref{eq:hamquadr}, re-arranging terms and using \eqref{eq:hamquadr} again, we get
\begin{equation}\label{eq:Ccons2}
  0 = \pi_a \Bigl(\frac{D\tilde{n}^a}{D\tau}
             -\frac{\pa\ham}{\pa \pi_a}(\tilde{n},\tilde{\pi})\Bigr)
      -n^a \Bigl(\frac{D\tilde{\pi}_a}{D\tau}
             +\frac{\pa\ham}{\pa n^a}(\tilde{n},\tilde{\pi})\Bigr)\;.
\end{equation}
Since $C$ should be conserved at any phase-space point $(n,\,\pi)$, we obtain that $\tilde{n}^a(\tau)$ and $\tilde{\pi}_a(\tau)$ must satisfy the Hamilton equations and, thus, the observable $C$ has to be of the form $C=W_{\tilde{n},\tilde{\pi}}$.

A similar statement can be obviously formulated in the configuration language: any conserved quantity linear in the trajectory $n(\tau)$ and its time derivative has to have the form \eqref{eq:consWr} for a solution $\tilde{n}(\tau)$.

Finally, thanks to the linear structure of the linearized system, the Wronskian of two solutions $n_1(\tau)$, $n_2(\tau)$ can be related to the Poisson bracket of the corresponding conserved quantities $W_{n_1,\pi_1}$ and $W_{n_2,\pi_2}$,
\begin{equation}\label{eq:PBWronskian}
    \{W_{n_1,\pi_1},W_{n_2,\pi_2}\} = W[n_1,\pi_1|n_2,\pi_2] \;.
\end{equation}

Let us finally return back to the geodesic motion and the corresponding linearized Jacobi system. In this case the Wronskian  \eqref{eq:Wronskiandef} takes the particular form
\begin{equation}\label{eq:WronskianJac}
    W[n_1|n_2] = m\Bigl( n_1^a \bar{g}_{ab} \frac{D n_2^b}{D\tau}
     - \frac{D n_1^a}{D\tau} \bar{g}_{ab} n_2^b\Bigr)\;,
\end{equation}
and the above formulae directly apply. In what follows we concentrate on this case.

\subsection{Canonical observables}

It is easy to see that the Poisson bracket of two observables linear in $(n,\pi)$ is equal to a constant, i.e., an observable independent of $(n,\pi)$. This observation can be used to construct a set of (time dependent) observables for the Jacobi system $(F^j,\,G_j)$, $j=1,\dots,n$, which form canonical coordinates at all times,
\begin{equation}\label{eq:cancoordFG}
    \{F^i,G_j\} = \delta_j^i\;,\quad
    \{F^i,F^j\} = \{G_i,G_j\} = 0\;.
\end{equation}
For example, choosing at time $\tau_0$ an orthonormal frame of vectors $e_{(i)}$ and the dual frame of 1-forms $e^{(i)}$, both at $\bar{x}(\tau_0)$, one can define the solutions of the Jacobi system $f^j(\tau)$ and $g_j(\tau)$ with initial values at $\tau_0$ given by
\begin{equation}\label{eq:fgdef}
\begin{aligned}
    f^j(\tau_0) &= e_{(j)}\;, &  \frac{D f^j}{D\tau}(\tau_0) &= 0 \;,\\
    g_j(\tau_0) &= 0\;, &  \frac{D g_j}{D\tau}(\tau_0) &= e^{(j)} \;.
\end{aligned}
\end{equation}
Clearly,
\begin{equation}\label{eq:fgWron}
    W[f^i,g_j]= \delta_j^i\;,\quad
    W[f^i,f^j]=W[g_i,g_j]=0
\end{equation}
at time $\tau_0$, and since the Wronskian is conserved, the relation \eqref{eq:fgWron} remains true at all times. Using this solution, and thanks to \eqref{eq:PBWronskian}, we can thus define canonical coordinates $(F^j,G_i)$ satisfying \eqref{eq:cancoordFG} by the corresponding Wronskian observables
\begin{equation}\label{eq:FGdef}
    F^j = W_{f^j}\;,\quad G_i = W_{g_i}\;.
\end{equation}
In particular, the set of coordinates $F^j$ (as well as the set of $G_i$) forms a maximal set of commuting conserved quantities of the linearized system. However, these conserved quantities are not very useful. To find them, one has to find first the solutions $f^j$ and $g_j$, i.e., to solve the linearized system.

In the following, we want to discuss more useful conserved quantities---given by the symmetries of the spacetime. To find these observables, in addition to symmetries one only needs to know the central trajectory.

\subsection{Conserved quantities generated by Killing tensors}

As we reviewed in Sec.~\ref{sec:geodesic}, generic (homogeneous in momentum) integrals of geodesic motion are generated by Killing tensors, see \eqref{eq:Killing_charge}. As shown by Caviglia, Zordan and Salmistraro (CZS) \cite{Caviglia:1982a,Caviglia:1982b} these tensors also generate the following linearized solutions $\tilde{n}^a(\tau)$, $\tilde{\pi}_a(\tau)$ for the Jacobi system:
\begin{align}
\tilde{n}^a &= \frac{\pa K}{\pa p_a}(\bar{x},\bar{p})
   = \frac{1}{(r{-}1)!} K^{a b_2 \dots b_{r}}(\bar{x})\, \bar{p}_{b_2} \dots \bar{p}_{b_{r}} \, ,\notag\\[-1.5ex]
   \label{eq:CZS}\\[-1.5ex]
\tilde{\pi}_a &= -\frac{\nabla_{\!a} K}{\pa x}(\bar{x},\bar{p})
   = - \frac{1}{r!} \nabla_{\!a}K^{b_1 \dots b_{r}}(\bar{x})\, \bar{p}_{b_1} \dots \bar{p}_{b_{r}} \, .\notag
\end{align}
Here, $K$ is the conserved quantity of geodesic motion generated by the Killing tensor $K^{a_1\dots a_r}$, \eqref{eq:Killing_charge}, $\frac{\pa K}{\pa p}$ denotes the derivative with respect to momentum $p$ with $x$ fixed, and $n^a \frac{\nabla_{\!a} K}{\pa x}(x,p)$ is the covariant derivative in direction $n^a$ with $p$ parallelly transported, {cf.~\eqref{covdobs} in Appendix~\ref{apx:PhsSpc}.} The latter derivative acts only on \mbox{$x$-dependent} terms in $K$ and essentially ignores momentum~$p$.

Let us verify the solution \eqref{eq:CZS} by checking the Hamilton equations \eqref{eq:covHE}. Taking advantage of the fact that the central trajectory is geodesic, $\frac{D\bar{p}_a}{D\tau}=0$, and $\bar{p}_a = m \bar{u}_a$, we obtain
\begin{equation}\label{eq:ntilHE1}
    \frac{D \tilde{n}^a}{D\tau} = \frac1{m}\frac{1}{(r{-}1)!} \bar{p}_{b_1} \nabla^{b_1} K^{a b_2 \dots b_{r}}(\bar{x})\,
    \bar{p}_{b_2} \dots \bar{p}_{b_{r}}\,.
\end{equation}
Using the identity
\be \label{eq:Killing_condition}
\nabla^a K^{b_1 \dots b_{r}} = - r  \nabla^{(b_1} K^{|a|b_2 \dots b_{r})}\;,
\ee
which follows from the Killing condition \eqref{eq:KillingTens},  yields
\begin{equation}\label{eq:ntilHE2}
\begin{split}
    \frac{D \tilde{n}^a}{D\tau}
    &= - \frac1{m}\frac{1}{r!} \nabla^{a} K^{b_1 \dots b_{r}}(\bar{x})\, \bar{p}_{b_1} \dots \bar{p}_{b_{r}} \\
    &= -\frac1{m} \frac{\nabla^{a} K}{\pa x} = \frac1{m} \bar{g}^{ab} \tilde{\pi}_b\;.
\end{split}
\end{equation}
For momentum $\tilde{\pi}_a$ we get
\begin{equation}\label{eq:pitilHE1}
\begin{split}
    \frac{D\tilde{\pi}_a}{D \tau}
   &= -\frac{1}{r!} \bar{u}^{c}\nabla_{\!c}\nabla_{\!a}K^{b_1 \dots b_{r}}(\bar{x})\,
     \bar{p}_{b_1} \dots \bar{p}_{b_{r}} \\
   &= -\frac1{m}\frac{1}{r!}\nabla_{\!a}\nabla^{(b_0}K^{b_1\dots b_r)}(\bar{x})\,
     \bar{p}_{b_0} \dots \bar{p}_{b_{r}} \\
     &\quad -\frac{1}{(r{-}1)!} \bar{u}^c \bar{R}_{ca}{}^{b_1}{}_{b}K^{bb_2\dots b_r}(\bar{x})\,
     \bar{p}_{b_1} \dots \bar{p}_{b_{r}}
     \, .
\end{split}\raisetag{9.7ex}
\end{equation}
Here, we have used the Ricci identity and the fact that all $r$ terms with the Riemann tensor give the same contribution. Now, the first term vanishes thanks to \eqref{eq:KillingTens} and in the second term we can identify $\tilde{n}^a$,
\begin{equation}\label{eq:pitilHE2}
\begin{split}
    \frac{D\tilde{\pi}_a}{D \tau}
     &= -m \bar{u}^c\bar{u}^d \bar{R}_{cadb}\frac{1}{(r{-}1)!}K^{bb_2\dots b_r}(\bar{x})\,
     \bar{p}_{b_2} \dots \bar{p}_{b_{r}}\\
     &=-m \bar{u}^c\bar{u}^d \bar{R}_{cadb}\tilde{n}^b
     \, ,
\end{split}\raisetag{4ex}
\end{equation}
which concludes the proof.

The CZS solution \eqref{eq:CZS} is of geometrical nature. It is obtained from the canonical transformation associated with the hidden symmetry of the geodesic equation. Let
{$\{\cdot,\cdot\}_g$}
be the Poisson bracket of the full geodesic theory, cf.~\eqref{eq:covPB}. Then $\tilde{n}^a = \{ x^a , K \}_g$ and $\tilde{\pi}_a=\{p_a,K\}_g$. This is the same as the infinitesimal transformation $\delta  x^a$, $\delta p_a$ of the central geodesic generated by the canonical transformation induced by $K$.

The solution \eqref{eq:CZS} can be related to the linearization $\mathscr{k}$ of the conserved quantity $K$. The expansion of any phase-space observable  $K(x,p)$ to the first order can be written as
\begin{equation}\label{eq:Kexp}
    K = \bar{K} + n^a \frac{\nabla_{\!a}K}{\pa x} + \pi_a \frac{\pa K}{\pa p_a}+\dots\;.
\end{equation}
Obviously, using \eqref{eq:CZS} we get
\begin{equation}\label{eq:Klin}
\begin{split}
    \mathscr{k} \equiv K- \bar{K}
      = \tilde{n}^a \pi_a - \tilde{\pi}_a n^a= W_{\tilde{n},\tilde{\pi}}\;.
\end{split}
\end{equation}
The linearized observable $\mathscr{k}=W_{\tilde{n},\tilde{\pi}}$ is thus again a conserved quantity that is linear in position and momentum and generated by the CZS solution $(\tilde{n},\,\tilde{\pi})$.

Let us finally mention that the CZS solution for $\tilde{n}^a$, \eqref{eq:CZS}, need not be generated from a Killing tensor. As noted in \cite{Caviglia:1982a, Caviglia:1982b} (see also \cite{Houri:2015lma,Cook:2009ef}), its existence is in one-to-one correspondence with a new object, called the affine tensor. An affine tensor of rank $r$, $K^{a_1\dots a_r}=K^{(a_1\dots a_r)}$,  is an object that satisfies
\be\label{Affine}
\nabla_{\!(a}K_{a_1\dots a_r)}=h_{aa_1\dots a_r}\,,\quad \nabla_{\!b} h_{aa_1\dots a_r}=0\,.
\ee
That is, the definition of an affine tensor `generalizes' that of a Killing tensor by requiring that its symmetrized derivative need not vanish but can be a covariantly constant tensor. Of course, the above presented construction of conserved quantities through Wronskians immediately generalizes to the CZS solutions generated by affine tensors. Let us stress, however, that although the Killing tensors are formally a subfamily of affine tensors, the requirement on the existence of non-trivial $h_{aa_1\dots a_r}$ is very strong and at the moment there are no known physical spacetimes admitting affine tensors that are not at the same time Killing tensors. For this reason we shall not probe this possibility in this paper any further.

\subsection{Integrability of the linearized system}

Let us now assume that we have at least two Killing tensors $K_1^{a\dots }$ and $K_2^{a\dots }$ corresponding to the conserved quantities $K_1$ and $K_2$ of the full geodesic motion.  In general, such integrals of motion do not Poisson-commute. Their Poisson bracket generates a new conserved quantity~$K$,
\begin{equation}\label{eq:PBKK}
    K = \{K_1, K_2\}_g\;,
\end{equation}
which corresponds also to a Killing tensor $K^{a\dots}$, given by the (symmetric) Schouten--Nijenhuis bracket \cite{Schouten:1940,Schouten:1954,Nijenhuis:1955}
\begin{equation}\label{eq:SNKK}
    K = [ K_1, K_2 ]_{\scriptscriptstyle\mathrm{SN}}\;,
\end{equation}
cf., e.g., \cite{FrolovKrtousKubiznak:2017review}.

For the linearized quantities $\mathscr{k}_1$, $\mathscr{k}_2$ we have
\begin{equation}\label{eq:PBlinkk1}
    \{\mathscr{k}_1,\mathscr{k}_2\}
      = \{W_{\tilde{n}_1,\tilde{\pi}_1},W_{\tilde{n}_2,\tilde{\pi}_2}\}
      = W[\tilde{n}_1,\tilde{\pi}_1|\tilde{n}_2,\tilde{\pi}_2]\;.
\end{equation}
The Wronskian can be expressed using the quantities related to the central trajectory. Substituting \eqref{eq:WronskianJacnpi} and \eqref{eq:CZS}, we get
\begin{equation}\label{eq:PBlinkk2}
\begin{split}
    \{\mathscr{k}_1,\mathscr{k}_2\}
      &=\Bigl(-\frac{\pa K_1}{\pa p_a}\frac{\nabla_{\!a}K_2}{\pa x} +
      \frac{\nabla_{\!a}K_1}{\pa x} \frac{\pa K_2}{\pa p_a}\Bigr)\bigg|_{\bar{x},\bar{p}}\\
      &=\{K_1,K_2\}_g\big|_{\bar{x},\bar{p}}=\bar{K}\;,
\end{split}
\end{equation}
the result already shown in \cite{Caviglia:1982b}.

It follows that if the original integrals of motion $K_1$, $K_2$ Poisson-commute, $K=0$, the linearized conserved quantities $\mathscr{k}_1$, $\mathscr{k}_2$ also Poisson-commute. In particular, if the spacetime geometry possesses a full set of commuting integrals of motion $K_j$ generated by Killing tensors, the linearized system has also a full set of mutually commuting integrals of motion $\mathscr{k}_j\equiv W_{\tilde{n}_j,\tilde{\pi}_j}$. The complete integrability of the geodesic motion thus naturally implies the complete integrability of the Jacobi system.

\section{Integrability of Jacobi equation in rotating black hole spacetimes\label{sec:black_holes}}

Let us now apply the above developed formalism to explicitly demonstrate the integrability of the Jacobi equation in the Kerr black hole spacetime \cite{Kerr:1963} and its higher-dimensional generalizations. Such an integrability stems from the existence of hidden symmetries in these spacetimes and derives from the integrability of the full geodesic motion. The same results remain also true for charged black holes and will be discussed elsewhere.

The Kerr metric represents a unique rotating black hole solution of vacuum Einstein equations.
In the Boyer--Lindquist coordinates it reads
\ba
ds^2 &=& - \frac{\Delta}{\Sigma} \left(dt - a \sin^2\!\theta d\varphi \right)^2 +
\frac{\sin^2\! \theta}{\Sigma} \left[ (r^2\! +\! a^2) d\varphi - a dt \right]^2  \nn \\
&& + \frac{\Sigma}{\Delta} dr^2 + \Sigma \, d\theta^2  \, , \label{eq:Kerr}
\ea
where $\Sigma = r^2 + a^2 \cos^2\!\theta$, and $\Delta = r^2 - 2Mr + a^2$.

The metric admits two Killing vectors $\pa_t$ and $\pa_\varphi$. For  a geodesic motion, these imply the following conserved charges:
\begin{align}
K_E &= - p_t  \label{eq:Kerr_charge_1}
\, , \\
K_L &= p_\varphi \label{eq:Kerr_charge_2}
\, .
\end{align}
In addition, there is a hidden symmetry encoded in the Killing tensor $K_{ab}$ that gives rise to Carter's constant \cite{Carter:1968rr,Carter:1968cmp,Walker:1970un}
\ba
\hspace{-0.25cm} K_{C} &=& \frac{1}{2} K^{ab} p_a p_b =  \frac{1}{2\Sigma} \left[ - \Delta  a^2 \cos^2\!\theta \, p_r^2  + r^2 p_\theta^2 \right. \nn \\
&&
\left. + \frac{a^2  \cos^2\!\theta}{\Delta} \left( (r^2+a^2)p_t  + a p_\varphi \right)^2  \right. \nn \\
&& \left. + \frac{r^2}{\sin^2\!\theta} \left( a \sin^2\!\theta \, p_t + p_\varphi \right)^2 \right]\,. \label{eq:Kerr_charge_3}
\ea
Lastly there is a conserved quantity generated by the metric $g_{ab}$, seen as a (covariantly constant) Killing tensor,
\ba
K_{m^2} &=& - g^{ab} p_a p_b = - \frac{1}{\Sigma} \left[ \Delta p_r^2 + p_\theta^2 + \frac{\Delta - a^2 \sin^2\!\theta}{\Delta \sin^2\!\theta} p_\varphi^2 \right. \nn \\
&& \left. - \frac{4 M a r}{\Delta} p_t p_\varphi - \frac{(r^2+a^2)^2 - a^2 \Delta \sin^2\!\theta}{\Delta} p_t^2 \right]\,.\quad \label{eq:Kerr_charge_4}
\ea
The four integrals of motion, $\{K_E, K_L, K_C, K_{m^2}\}$, are functionally independent and mutually Poisson commute, yielding the geodesic motion completely integrable \cite{Carter:1968rr}.
The explicit solution in terms of special functions can be for example found in \cite{Hackmann:2009nh, Hackmann:2010zz}.

We can now pick our favourite geodesic and turn to the corresponding linearized Jacobi system.
The CZS solution, \eqref{eq:CZS}, yields the following independent solutions:
\ba
\tilde{n}_E &=& - \pa_t \, , \\
\tilde{n}_L &=&   \pa_\varphi \, ,  \\
\Sigma \tilde{n}_{C} &=&  - \Delta a^2 \cos^2\!\bar{\theta} \bar{p}_r \pa_r + \bar{r}^2 \bar{p}_\theta \pa_\theta \nn \\
&& + \left( \frac{\bar{p}_\varphi}{\sin^2\!\bar{\theta}} + a \bar{p}_t \right) \bar{r}^2  \pa_\varphi \nn  \\
&& \hspace{0cm}  +
  \frac{a^3 \cos^2\!\bar{\theta} \left( a \bar{p}_\varphi + \bar{p}_t (\bar{r}^2 + a^2)\right)}{\Delta} \pa_\varphi  \nn \\
&&   + \frac{a^2 \cos^2\!\bar{\theta} (\bar{r}^2 + a^2)\left( a \bar{p}_\varphi + (\bar{r}^2+a^2)\bar{p}_t \right)}{\Delta} \pa_t  \nn \\
&&   + a \bar{r}^2 ( \bar{p}_\varphi + a \sin^2\!\bar{\theta} \bar{p}_t)  \pa_t  \, ,  \\
- \frac{\Sigma}{2} \tilde{n}_{m^2} &=&  \Delta \bar{p}_r \pa_r + \bar{p}_\theta \pa_\theta \nn \\
&& + \left( \frac{\Delta - a^2 \sin^2\!\bar{\theta}}{\Delta \sin^2\!\bar{\theta}} \bar{p}_\varphi - \frac{2 M a r}{\Delta} \bar{p}_t \right) \pa_\varphi   \nn \\
&& \hspace{-1,25cm} - \left( \frac{(\bar{r}^2+a^2)^2 - a^2 \Delta \sin^2\!\bar{\theta}}{\Delta} \bar{p}_t + \frac{2 M a r}{\Delta}  \bar{p}_\varphi \right) \pa_t\,.
\ea
From these we construct the independent conserved quantities for the Jacobi equation in Kerr
\be \label{eq:Kerr_conserved_quantities}
\mathscr{k}_i = \tilde{n}_{i}^a \pi_a - n^a \frac{D\tilde{n}_{i a}}{D\tau} \, , \qquad i = 1, \dots , 4 \, .
\ee
We set the constant values of the Wronskians to $w_i$, $i=1, \dots, 4$, and introduce the abbreviated notation $g_i = n^a \frac{D\tilde{n}_{i a}}{D\tau}$, where the $g_i$ do not depend on the momenta $\pi$. From \eqref{eq:Kerr_conserved_quantities} it is easy to extract the value of the momenta $\pi_t$ and $\pi_\varphi$, which are given by
\ba
\pi_t = \pi_t (\bar{x}^a, n^a) &=& - g_1 - w_1 \, , \\
\pi_\varphi = \pi_\varphi  (\bar{t},\bar{r}, \bar{\theta}, \bar{\varphi}) &=& g_2 + w_2 \, .
\ea
For geodesics with $p_r \neq 0$, $p_\theta \neq 0$ it is possible to invert \eqref{eq:Kerr_conserved_quantities} with respect to $\pi_r$, $\pi_\theta$.  The result is
\ba
\pi_r &=& \alpha_r + \beta_r \pi_\varphi + \gamma_r \pi_t \, ,\\
\pi_\theta &=& \alpha_\theta + \beta_\theta \pi_\varphi + \gamma_\theta \pi_t \, ,
\ea
with
\ba
\alpha_r &=& - \frac{\bar{r}^2 f_3 + 2 f_4}{2 \Delta p_r} \, ,  \\
\beta_r &=& \frac{a}{\Delta^2 p_r} \left[ (a^2+\bar{r}^2) p_t + a p_\varphi \right] \, ,  \\
\gamma_r &=& \frac{a^2+\bar{r}^2}{a} \beta_r \, ,\\
\alpha_\theta &=&  \frac{1}{2 p_\theta} \left( - a^2 \cos^2\bar{\theta} f_3 + 2 f_4 \right)   \, ,  \\
\beta_\theta &=&  - \frac{1}{p_\theta} \left[  a  p_t + \frac{1}{\sin^2\bar{\theta}} p_\varphi \right] \, ,  \\
\gamma_\theta &=& a \sin^2\bar{\theta} \beta_\theta \,.
\ea
Here we have set $f_3 = w_3 + g_3$, $f_4 = w_4 + g_4$,  these are functions of $n^a$ and not of the momenta.
These expressions are involved, although in a closed form: it is a good example of the fact that the Jacobi equation is complicated even if it is linear.

Let us finally mention that the procedure described in this section directly generalizes to higher-dimensional Kerr-NUT-(A)dS black hole spacetimes \cite{Chen:2006xh}. Such spacetimes are known to admit a number of hidden symmetries that yield the geodesic motion completely integrable \cite{FrolovStojkovic:2003b, Page:2006ka}. It follows that the corresponding Jacobi system is also integrable and in principle can be solved by the same steps described in this section. Let us, however, stress that in higher dimensions, the generic geodesic is given only in terms of complicated integrals \cite{FrolovKrtousKubiznak:2017review}, see also  \cite{Hackmann:2008tu, Kagramanova:2012hw, Diemer:2014lba} for special cases, and the solution of the Jacobi system thus becomes far from explicit.

\section{Conclusions\label{sec:conclusions}}

In this paper we have analyzed the Jacobi geodesic deviation equation from a point of view of Hamiltonian dynamics. It represents a dynamical system that is (although linear) explicitly time dependent. Consequently, the coordinate Hamiltonian is not covariant and varies from chart to chart. Nevertheless, we have shown that a covariant Hamiltonian can be constructed and shown (see Appendix~\ref{apx:CovLin}) how it can be obtained by the canonical transformation (accompanied by a due linearization) of the geodesic Hamiltonian. Although the geodesic Hamiltonian is a constant of motion, the linearized Hamiltonian for the geodesic deviation depends explicitly on time.

The main result of our paper regards the observation that the integrals of geodesic motion give rise to the corresponding integrals for the Jacobi system that are linear and given by the invariant Wronskians. In particular, this is true for the integrals generated by hidden symmetries of the spacetime. We have shown that if the geodesic motion is completely integrable, so will be the corresponding linearized motion described by the Jacobi equation. This has been further illustrated on an example of rotating black hole spacetimes in four and higher dimensions.

There is a number of topics that we have not discussed and that we point out as suitable for future research. One of these is the inclusion of spin. For example, a Lagrangian for the Jacobi deviation in the presence of Grassmannian spin variables can be found in \cite{vanHolten:2001ea}, and a discussion of conserved quantities for geodesics in the presence of spin in Kerr--NUT--(A)dS spaces in \cite{Kubiznak:2011ay}. Another one is the possibility of extending our results to higher order geodesic perturbations. These have been used to build analytic approximations of generic geodesics from simple exact solutions  \cite{Kerner:2001cw,Colistete:2002ka}, and have been used to model extreme mass-ratio systems \cite{Koekoek:2011mm}. It would be interesting to find out if for example the standard conserved charges of Kerr can be used to build conserved charges for higher order geodesic perturbations. Lastly our results can be used to discuss the issue of (in)stability of dynamical  systems \cite{Casetti:2000gd,Pettini:book2007}.

\vspace{0.2cm}

\begin{acknowledgements}
M.C.\ and T.H., who met for the first time at the 2nd workshop on ``Quantum Aspects of Black Holes and its Recent Progress'' held in Yerevan, Armenia and then started this project, would like to thank the organizers of the workshop for providing this stimulating opportunity. M.C. acknowledges CNPq support from projects (205029/2014-0) and (303923/2015-6), and a Pesquisador Mineiro project n.~PPM-00630-17. M.C. would like to thank the Physics Department at the University of Padova, Italy, and the Institute of Theoretical Physics of the Charles University, Czech Republic, for hospitality during the initial stages of the project. P.K.\ was supported by the grant No.~14-37086G of the Czech Science Foundation. D.K.\ was supported by the Perimeter Institute for Theoretical Physics and by the Natural Sciences and Engineering Research Council of Canada. Research at Perimeter Institute is supported by the Government of Canada through
the Department of Innovation, Science and Economic Development Canada and by
the Province of Ontario through the Ministry of Research, Innovation and Science.
\end{acknowledgements}

\appendix

\section{Jacobi equation: phase space formalism}
\label{apx:PhsSpc}

In this appendix we want to elucidate the origin of the Jacobi equation from the point of view of the phase space formalism. We start by discussing the  linearized evolution of a completely general system and re-derive the material from section \ref{sc:symdyn} for a general phase space, without the cotangent bundle structure. Specifying next to a phase space with the cotangent bundle structure and the Hamiltonian given by the configuration space metric, we show how to return back to the previous formalism.

\subsection{Linearized phase space and Wronskian}

In general, the phase space is a symplectic space with the symplectic structure $\Omega$. It allows to define the Poisson brackets of two observables\footnote{%
Overview of the phase space description can be found in many standard textbooks. We follow conventions summarized, for example, in the review \cite{FrolovKrtousKubiznak:2017review}. In particular, we use capital Latin letters ${\scriptstyle\pix{A}},\,{\scriptstyle\pix{B}},\,\dots$ as phase-space indices and we denote general coordinates on the phase space as $\mathcal{X}^{\pix{A}}$. The symplectic structure has components $\Omega_{\pix{AB}}$. Its inverse,  $\Omega^{\pix{AB}}$, satisfies $\Omega_{\pix{AC}}\Omega^{\pix{BC}}=\delta_{\pix{A}}^{\pix{B}}$. Occasionally, we skip tensor indices if the tensorial structure is clear from the context.}
\begin{equation}\label{PB}
    \{F,G\} = F_{,\pix{A}}\, \Omega^{\pix{AB}}\, G_{,\pix{B}}\;,
\end{equation}
and the Hamiltonian vector flow $X_{\!\obsix{F}}$ associated with an observable $F$
\begin{equation}\label{Hamvect}
    X_{\!\obsix{F}}^{\pix{A}} = \Omega^{\pix{AB}}\, F_{,\pix{B}}\;.
\end{equation}
Given a Hamiltonian $H$, the time evolution (classical trajectories) are given by orbits of the Hamiltonian flow~$X_{\!\obsix{H}}$.

Assuming such a congruence of classical phase-space trajectories generated by $X_{\!\obsix{H}}$, let us pick up one particular trajectory $\bar{\mathcal{X}}(\tau)$, which we call the central trajectory. We want to study a one-parametric subfamily $\mathcal{X}(\sigma,\tau)$ of these classical trajectories,
\begin{equation}\label{genevol}
    X_{\!\obsix{H}}^{\pix{A}} = \frac{d \mathcal{X}^{\pix{A}}}{d\tau}(\sigma,\tau)\;,
\end{equation}
which is close to the central trajectory, $\mathcal{X}(0,\tau)=\bar{\mathcal{X}}(\tau)$. In a linear approximation, such a family is generated by a phase-space vector field $N(\tau)$ along the central trajectory $\bar{\mathcal{X}}(\tau)$,
\begin{equation}\label{Ndef}
    N^{\pix{A}}(\tau) = \frac{d \mathcal{X}^{\pix{A}}}{d\sigma} (0,\tau)\;.
\end{equation}
We call $N(\tau)$ a linearized trajectory based on the central trajectory $\bar{\mathcal{X}}(\tau)$.

We can extend the definition of $N$, \eqref{Ndef}, also to nonzero values of $\sigma$. Since the vector fields $X_{\!\obsix{H}}$ and  $N$ can be viewed as coordinate fields $X_{\!\obsix{H}}\equiv\partial_\tau$ and $N\equiv\partial_\sigma$ on a two-dimensional sheet $\mathcal{X}(\sigma,\tau)$ with coordinates $\sigma$ and ${\tau}$, they must Lie-commute,
\begin{equation}\label{Liecom}
    \bigl[X_{\!\obsix{H}}, N\bigr] = 0\;.
\end{equation}
This equation can be also rephrased as a condition on $N(\tau)$ along the trajectory $\bar{\mathcal{X}}(\tau)$,
\begin{equation}\label{linphspeom}
    \mathcal{L}_{X_{\!\obsix{H}}} N = 0\;.
\end{equation}
Either of the last two equations can be viewed as the equation of motion for the linearized trajectory $N(\tau)$.

The space of phase-space vectors at $\bar{\mathcal{X}}(\tau)$ represents the linearized phase space at time $\tau$. It is a linear symplectic space with the symplectic structure given by the original symplectic structure $\Omega_{\pix{AB}}$ evaluated at $\bar{\mathcal{X}}(\tau)$. Similar to the previous discussion, where the phase space has been represented by pairs ${(n,\pi)}$, the linearized phase spaces at different times are geometrically different spaces. Therefore, we cannot directly apply the standard Hamiltonian formalism. We have seen that for that we would need to identify somehow the spaces at different times. However, let us discuss general phase-space situation without such an identification first. To this purpose we use the fact that the linearized evolution is given by the condition \eqref{linphspeom} which naturally relates the phase spaces at different times.

As we have seen, the linear structure of linearized phase spaces allows us to define the Wronskian of two phase-space vectors $N_1,\,N_2$. It is given by the symplectic structure as
\begin{equation}\label{Wronskian}
    W[N_1|N_2] = N_1^{\pix{A}} \, \Omega_{\pix{AB}} \,N_2^{\pix{B}}\;,
\end{equation}
cf.~\eqref{eq:WronskianJacnpi}.
It immediately follows that it is conserved along the time evolution
\begin{equation}\label{consWr}
\begin{split}
  &\frac{d}{d\tau} W[N_1|N_2] = \mathcal{L}_{X_{\!\obsix{H}}} W[N_1|N_2] \\
  &\quad= (\mathcal{L}_{X_{\!\obsix{H}}} N_1^{\pix{A}}) \,\Omega_{\pix{AB}}\,N_2^{\pix{B}}
    +N_1^{\pix{A}}\,\Omega_{\pix{AB}}\,(\mathcal{L}_{X_{\!\obsix{H}}} N_2^{\pix{B}}) = 0\;.
\end{split}\raisetag{6ex}
\end{equation}
Here, we have used \eqref{linphspeom} and the fact that the symplectic structure is conserved along any Hamiltonian flow, $\mathcal{L}_{X_{\!\obsix{F}}}\Omega = 0$. Any chosen linearized solution $\tilde{N}(\tau)$ thus defines a conserved quantity $W_{\!\obsix{\tilde{N}}}$ on linearized solutions,
\begin{equation}\label{linconsquant}
    W_{\!\obsix{\tilde{N}}}(N) = W[\tilde{N}|N]\;.
\end{equation}

\subsection{Conserved quantities and integrability}

Let us now assume that the original system admits an integral of motion $K$, i.e., that there exists an observable~$K$ (not explicitly dependent on time parameter $\tau$) which Poisson-commutes with the Hamiltonian,
\begin{equation}\label{consqK}
    \{K,H\}=0\;.
\end{equation}
A simple manipulation yields that
\begin{equation}\label{constimplsol}
    \mathcal{L}_{X_{\!\obsix{H}}} X_{\!\obsix{K}}=
    \bigl[X_{\!\obsix{H}},X_{\!\obsix{K}}\bigr]=X_{\!\textstyle \{K,H\}}=0\;.
\end{equation}
Any conserved quantity $K$ thus induces a linearized solution given by $X_{\!\obsix{K}}$ evaluated along $\bar{\mathcal{X}}(\tau)$. Clearly, the Wronskian of such two linearized trajectories is given by the Poisson bracket of the original conserved quantities,
\begin{equation}\label{WKK}
    W[X_{\!\obsix{K_1}}|X_{\!\obsix{K_2}}] = \{K_1,K_2\}\;.
\end{equation}

Following the previous discussion of the Jacobi system, a linearized solution $X_{\!\obsix{K}}$ associated with a conserved quantity $K$ defines a conserved quantity $\mathscr{k}$ on the linearized phase space by the relation \eqref{linconsquant} above,
\begin{equation}\label{kscrdef}
    \mathscr{k}(N) = W_{\!\obsix{X_{\!\obsix{K}}}}(N)\;.
\end{equation}
It is straightforward to show that
\begin{equation}\label{linerizationconsquant}
    \mathscr{k}(N) = W[X_{\!\obsix{K}}|N] = N^{\pix{A}} K_{,\pix{A}}\;,
\end{equation}
cf.\ eqs.~\eqref{Wronskian} and \eqref{Hamvect}. It means, that $\mathscr{k}$ is the linearization of the original conserved quantity $K$,
\begin{equation}\label{LKrel}
    K(\mathcal{X}(\sigma,\tau)) =
    K(\bar{\mathcal{X}}(\tau))  + \sigma\, \mathscr{k}(N(\tau))+\mathcal{O}(\sigma^2)\;,
\end{equation}
i.e., ${\mathscr{k}(\sigma N) = K-\bar{K}}$, cf.~\eqref{eq:Klin}.

Till now in this section, we have not assumed anything particular about the phase space and the Hamiltonian. We have just observed that any linearized solution defines through the Wronskian a conserved quantity \eqref{linconsquant} and that any conserved quantity of the original system defines the linearized solution $X_{\!\obsix{K}}$ and the linearized conserved quantity ${\mathscr{k}}$ which is explicitly linear in $N$, as seen from \eqref{linerizationconsquant}. We see that such a construction is completely general.

For a completely integrable system we have $n$ mutually Poisson-commuting integrals of motion $K_i$. They generate linearized solutions $X_{\!\obsix{K_i}}$. Thanks to the linearity of \eqref{linphspeom} any linear (with constant coefficients) combination $N$ of these solutions 
is again a linearized solution. Moreover, each of the conserved quantities $K_i$ induces the linearized quantity ${{\mathscr{k}}_i}$. Evaluating this linearized quantity on solutions $X_{\!\obsix{K_j}}$, one finds
\begin{equation}\label{LonKzero}
    {\mathscr{k}}_i(X_{\!\obsix{K_j}}) = \{K_i,K_j\} = 0\;.
\end{equation}
The same is true for any linear combination $N$ of $X_{\!\obsix{K_j}}$, i.e., ${{\mathscr{k}}_i}(N) = 0$. The conserved quantities $K_j$ thus directly generate a family of linearized solutions which all have vanishing values of linearized quantities ${\mathscr{k}}_i$. In other words, all these solutions have the same values of the conserved quantities $K_i$ as the central trajectory, ${K_j = \bar{K}_j}$.

It is not surprising since for the integrable system $X_{\!\obsix{K_j}}$ generate symmetries of the evolution. These vector fields are tangent to the Lagrangian submanifolds given by $K_j=\text{const}$. The linearized solutions $X_{\!\obsix{K_j}}$ thus corresponds to trajectories which remain in the same Lagrangian submanifold as the central trajectory.

To conclude, we have just seen that the linearized conserved quantities constructed by using Wronskian and the linearized solutions of the equations of motion generated by the integrals of motion of the full system are general features of any Hamiltonian system admitting integrals of motion.

\subsection{Cotangent bundle structure of the phase space}

Let us now relate this general formulation to the configuration space description presented in sections \ref{sc:JacobiEq} and \ref{sc:symdyn}. To that purpose we consider the phase space to be built from a configuration space $\mathcal{M}$. While the configuration space is a space of ``positions'' $x$, the phase space is a space of ``positions and momenta'' $(x,p)$. It is well known that such a phase space can be represented as a cotangent bundle $\mathbf{T}^* \mathcal{M}$. The cotangent bundle has a natural symplectic structure $\Omega$. If one chooses the configuration-space coordinates $x^a$ and the components $p_a$ of the momentum with respect to the frame $dx^a$ as coordinates in the phase space, $\mathcal{X}^{\pix{A}}\equiv(x^a, p_a)$, the symplectic structure takes the canonical form,
\begin{equation}\label{cancoorOmega}
    \Omega = dx^a \wedge dp_a\;,
\end{equation}
where the sum over spacetime index is naturally assumed. It means that $(x^a, p_a)$ are canonical coordinates in which the Poisson bracket takes the following form:
\begin{equation}\label{cancoorPB}
    \{F,G\} = \frac{\partial F}{\partial x^a}\frac{\partial G}{\partial p_a}
             -\frac{\partial F}{\partial p_a}\frac{\partial G}{\partial x^a}\;.
\end{equation}

A phase-space tangent vector $N$ can be written with respect to coordinate frame $\frac{\partial}{\partial x^a}$ and $\frac{\partial}{\partial p_a}$ as
\begin{equation}\label{Ncoordecomp}
    N = n^a \frac{\partial}{\partial x^a} + {\tilde\pi}_a \frac{\partial}{\partial p_a}\;.
\end{equation}
Components $n^a$ can be understood as components of a configuration-space vector $n$, which is independent of the choice of coordinates $x^a$. On other hand, components ${\tilde\pi}_a$ cannot be combined to a 1-form $\pi$ which would be independent of the choice of coordinates $x^a$. Splitting \eqref{Ncoordecomp} to the configuration and momentum part is coordinate dependent.

However, one can formulate a similar decomposition in a covariant way. As described in Appendix of \cite{Cariglia:2012qj}, having a (torsion-free) covariant derivative $\nabla$ on the configuration space, one can introduce the covariant splitting
\begin{equation}\label{Ncovdecomp}
    N = n^a \frac{\nabla_{\!a}}{\partial x} + \pi_a \frac{\partial}{\partial p_a}\;.
\end{equation}
Here, the first term corresponds to a direction in the phase space given by the changing position ${x\to x+\varepsilon n}$ with momentum $p$ covariantly fixed (i.e., parallelly transported along $n$ using the covariant derivative $\nabla$). The second term corresponds to a direction in the phase space given by the changing momentum ${p\to p+\varepsilon\pi}$ with $x$ fixed. Such a splitting uniquely relates a phase-space vector $N$ with a pair of the configuration-space vector $n$ and the configuration-space 1-form $\pi$. Phase-space vectors $\frac{\nabla_{\!a}}{\partial x}$ and $\frac{\partial}{\partial p_a}$ thus describe  horizontal and vertical directions in the cotangent bundle, where the `horizontality' is given by the covariant derivative $\nabla$.

Using this decomposition one can define in a covariant way derivatives of an observable $F$ with respect to the position and the momentum,
\begin{equation}\label{covdobs}
   \frac{\nabla_{\!a} F}{\partial x}\equiv \frac{\nabla_{\!a}^{\pix{A}}}{\partial x}\,F_{\!,\pix{A}}\;,\quad
   \frac{\partial F}{\partial p_a}\equiv \frac{\partial^{\pix{A}}}{\partial x^a} \,F_{\!,\pix{A}}\;.
\end{equation}
In terms of these, the Poisson bracket reads
\begin{equation}\label{eq:covPB}
    \{F,G\} = \frac{\nabla_{\!a} F}{\partial x}\frac{\partial G}{\partial p_a}
             -\frac{\partial F}{\partial p_a}\frac{\nabla_{\!a} G}{\partial x}\;.
\end{equation}

The linearized equations of motion can be written in the form \eqref{Liecom}. It will be useful to write the Lie bracket of two phase-space vector fields $N_1$, $N_2$ in terms of the covariant decomposition. In the splitting \eqref{Ncovdecomp} each vector field $N_i(x,p)$ corresponds to a pair $n_i(x,p)$, $\pi_i(x,p)$ of the configuration-space vector and \mbox{1-form}, which both depend  on $x$ and $p$. Acting with the Lie bracket on a phase-space scalar $F$, one gets
\begin{widetext}
\begin{equation}\label{Liebrdecomp}
\begin{aligned}
  &\Bigl[N_1,\,N_2\Bigr]F
    =n_1^a\,n_2^b\Bigl[
      \frac{\nabla_{\!a}}{\partial x}\frac{\nabla_{\!b}}{\partial x}
      -\frac{\nabla_{\!b}}{\partial x}\frac{\nabla_{\!a}}{\partial x}
      \Bigr]F
    +\pi_{1a}\pi_{2b}\Bigl[
      \frac{\partial}{\partial p_a}\frac{\partial}{\partial p_b}
      -\frac{\partial}{\partial p_b}\frac{\partial}{\partial p_a}
      \Bigr]F
    +\bigl(n_1^a\,\pi_{2b}-n_2^a\,\pi_{1b}\bigr)\Bigl[
      \frac{\nabla_{\!a}}{\partial x}\frac{\partial}{\partial p_b}
      -\frac{\partial}{\partial p_b}\frac{\nabla_{\!a}}{\partial x}
      \Bigr]F
    \\&\quad
    +\Biggl(n_1^a\frac{\nabla{}^{2\!}n^b}{\partial x}
    -n_2^a\frac{\nabla{}^{1\!}n^b}{\partial x}
    +\pi_{1a}\frac{\partial {}^{2\!}n^b}{\partial p_a}
    -\pi_{2a}\frac{\partial {}^{1\!}n^b}{\partial p_a}\Biggr)
      \frac{\nabla_{\!b} F}{\partial x}
    +\Biggl(n_1^a\frac{\nabla\pi_{2b}}{\partial x}
    -n_2^a\frac{\nabla{}^{1\!}\pi_b}{\partial x}
    +\pi_{1a}\frac{\partial \pi_{2b}}{\partial p_a}
    -\pi_{2a}\frac{\partial {}^{1\!}\pi_b}{\partial p_a}\Biggr)
      \frac{\partial F}{\partial p_b}\;.
\end{aligned}
\end{equation}
It can be shown that the first term is non-trivial while the next two terms vanish,
\begin{equation}\label{xxderterm}
    \Bigl[
      \frac{\nabla_{\!a}}{\partial x}\frac{\nabla_{\!b}}{\partial x}
      -\frac{\nabla_{\!b}}{\partial x}\frac{\nabla_{\!a}}{\partial x}
      \Bigr]F =  p_k\, R{}_{ab}{}^k{}_l\, \frac{\partial F}{\partial p_l}\;,\qquad
  \Bigl[
      \frac{\partial}{\partial p_a}\frac{\partial}{\partial p_b}
      -\frac{\partial}{\partial p_b}\frac{\partial}{\partial p_a}
      \Bigr]F = 0\;,\qquad
  \Bigl[
      \frac{\nabla_{\!a}}{\partial x}\frac{\partial}{\partial p_b}
      -\frac{\partial}{\partial p_b}\frac{\nabla_{\!a}}{\partial x}
      \Bigr]F =0\;,
\end{equation}
see the end of this Appendix. The covariant splitting of the Lie bracket thus reads
\begin{equation}\label{Liebrdecomp}
\begin{aligned}
  \Bigl[N_1,\,N_2\Bigr]
    &=
    \Biggl(n_1^a\frac{\nabla{}^{2\!}n^b}{\partial x}
    -n_2^a\frac{\nabla{}^{1\!}n^b}{\partial x}
    +\pi_{1a}\frac{\partial n_2^b}{\partial p_a}
    -\pi_{2a}\frac{\partial n_1^b}{\partial p_a}\Biggr)
      \frac{\nabla_{\!b}}{\partial x}
    \\&
    +\Biggl(n_1^a\frac{\nabla\pi_{2b}}{\partial x}
    -n_2^a\frac{\nabla\pi_{1b}}{\partial x}
    +\pi_{1a}\frac{\partial \pi_{2b}}{\partial p_a}
    -\pi_{2a}\frac{\partial \pi_{1b}}{\partial p_a}
    +n_1^c\,n_2^d\, p_a\, R{}_{cd}{}^a{}_b\Biggr)
      \frac{\partial}{\partial p_b}\;.
\end{aligned}
\end{equation}
\end{widetext}

Let us now turn to the study of a geodesic motion in the configuration space. We assume the existence of a metric $g_{ab}$ and naturally choose $\nabla$ to be the metric (torsion-free) covariant derivative, $\nabla_{\!c} g_{ab}=0$. The geodesic motion in the configuration space is given by a simple quadratic Hamiltonian
\begin{equation}\label{geodHamiltonian}
    H(x,p) = \frac1{2m} g^{ab}(x)\, p_a p_b\;.
\end{equation}
Since the metric is covariantly constant, we have $\frac{\nabla_{\!a} H}{\partial x} = 0$ and $\frac{\partial H}{\partial p_a} = g^{ab} p_b$. Therefore, the covariant splitting of the Hamiltonian flow is
\begin{equation}\label{geodHamflow}
    X_{\!\obsix{H}} = \frac{\partial H}{\partial p_a}\frac{\nabla_{\!a}}{\partial x}
     - \frac{\nabla_{\!a} H}{\partial x}\frac{\partial}{\partial p_a}
     = \frac1m\,p^a \frac{\nabla_{\!a}}{\partial x}\;.
\end{equation}

Now, let us return to the linearized equations of motion \eqref{linphspeom} near the central geodesic $\bar{\mathcal{X}}(\tau)$, which is given in the configuration-space language by $\bar{x}(\tau)$ and $\bar{p}(\tau)$. The linearized trajectory $N$ is given by \eqref{Ncovdecomp} and $X_{\!\obsix{H}}$ by  \eqref{geodHamflow}. Substituting these into \eqref{linphspeom} and employing \eqref{Liebrdecomp}, the linearized equations of motion  yield
\begin{equation}\label{eomsplit}
\begin{split}
  &\mathcal{L}_{X_{\!\obsix{H}}} N = \bigl[X_{\!\obsix{H}},N\bigr]
    = \frac1m\Bigl(\bar{p}^a\frac{\nabla_{\!a}n^b}{\partial x}
      - \pi_a \bar{g}^{ab}\Bigr)\frac{\nabla_{\!b}}{\partial x}
  \\&\mspace{50mu}
    + \frac1m\Bigl(\bar{p}^a\frac{\nabla_{\!a}\pi_b}{\partial x}
      + \bar{p}_c \bar{p}_d\,\bar{R}{}^c{}_b{}^d{}_a\, n^a \Bigr)
      \frac{\partial}{\partial p_b} =0\;.
\end{split}
\end{equation}
Here, as before, the bar indicates quantities evaluated at the central geodesic. For the Hamiltonian \eqref{geodHamiltonian}, the momentum $\bar{p}$ of the central geodesic is proportional to the velocity, $\bar{p}=m\bar{u}$, and therefore $\frac1m\bar{p}^a\frac{\nabla_{\!a}}{\partial x}\equiv \frac{\nabla}{d\tau}$ is the covariant time derivative along the central geodesic. We finally obtain \eqref{eq:covHE},
\begin{equation}\label{JacobiHamEq}
    \frac{D n^a}{D\tau} = \frac1m\, \pi^a\;,\qquad
    \frac{D \pi_a}{D\tau} = m\, \bar{u}^k \bar{u}^l\,\bar{R}_{aklb}\, n^b\;,
\end{equation}
which gives the Jacobi equation for the linearized geodesics,
\begin{equation}\label{JacobiHamEq}
    \frac{D^2 n^a}{D\tau^2} = \bar{R}^a{}_{cdb}\,\bar{u}^c \bar{u}^d\, n^b\;.
\end{equation}

\subsection{Commutation relations for derivatives on the cotangent bundle}

To prove the relations \eqref{xxderterm}, it is useful to consider an observable $F$ monomial in momentum,
\begin{equation}\label{eq:monobs}
 F(x,p)=\frac{1}{r!}\,f^{c_1\dots c_r}(x)\,p_{c_1}\dots p_{c_r}\;,
\end{equation}
where $f^{c_1\dots c_r}$ is a symmetric configuration-space tensor.
A generic observable can be then obtained as a sum of monomial observables.

The derivative of such an observable with respect to the momentum is
\begin{equation}\label{eq:mommonoobs}
    \frac{\pa F}{\pa p_a} = \frac{1}{(r{-}1)!}\, f^{ac_2\dots c_r}\, p_{c_2}\dots p_{c_r}\;.
\end{equation}
The commutator of the covariant derivatives with respect to the position reads
\begin{equation}\label{eq:poscommonoobs}
\begin{split}
    2\frac{\nabla_{\![a}}{\pa x}\frac{\nabla_{\!b]}}{\pa x} F
    &= \frac{2}{r!}\, \nabla_{\![a}\nabla_{\!b]}f^{c_1\dots c_r}\, p_{c_1}\dots p_{c_r}\\
    &= \frac{r}{r!}R_{ab}{}^{c_1}{}_d \, f^{dc_2\dots c_r}\, p_{c_1} p_{c_2}\dots p_{c_r}\\
    &= p_c \, R_{ab}{}^{c}{}_{d} \,\frac{\pa F}{\pa p_d}
    \;,
\end{split}
\end{equation}
where in the first step we have used the fact that all $r$ contributions from the Ricci identities are the same and in the second step employed the relation \eqref{eq:mommonoobs}.

The remaining two commutators in \eqref{xxderterm} are trivial. They reflect the fact that there is no curvature in $p$ directions.

\section{Covariant Lagrangian and Hamiltonian for the Jacobi system}
\label{apx:CovLin}

\subsection{Covariant expansion of the Lagrangian}

The dynamics of a linearized system, i.e., the expansion of the equations of motion to the first order in the deviation variable, can be derived from the approximation of the Lagrangian to the second order. The first order contribution to the Lagrangian is trivial (given by a total derivative) since we are expanding near a solution of the equations of motion, i.e., near an extremal trajectory.

In order to expand the Lagrangian to the second order, we have to first generalize the notion of a deviation vector. In the main text, we have introduced the deviation vector ${n^a}$ as a tangent vector in \mbox{${\sigma}$-direction} of the family of trajectories ${x(\sigma,\tau)}$. This approach is sufficient for the first order approximation, e.g., for the derivation of the Jacobi equation \eqref{eq:Jacobi}. However, for an approximation to a higher order, and in particular to derive the Lagrangian \eqref{eq:Lagrangian_vanHolten}, one has to first define the deviation vector more carefully.

We start again by choosing the central geodesic $\bar{x}(\tau)$ parameterized by its proper time $\tau$ and use this time as the external time for general trajectory $x(\tau)$ nearby. Next we perform a time dependent change of the configuration  variables, where instead of a configuration point $x$ at time $\tau$ we use the deviation vector $n^a$ at $\bar{x}(\tau)$,
\begin{equation}\label{eq:nxtransf}
    n^a = n^a(\bar{x}|x)\;.
\end{equation}
That is, we define ${n^a(\bar{x}|x)}$ as a vector at ${\bar{x}}$ tangent to the geodesic joining points ${\bar{x}}$ and ${x}$, assuming that ${x}$ is in a normal neighborhood of ${\bar{x}}$. To specify this vector uniquely, we normalize it to the geodesic distance of the corresponding geodesic segment. Intuitively, this vector plays a role of a difference between positions, ${n=x-\bar{x}}$, generalized to the curved spacetime.

We stress that the transformation \eqref{eq:nxtransf} is time dependent through the dependence on the point $\bar{x}$ on the central geodesic. (For brevity, in what follows we will not write this $\tau$-dependence explicitly.) This has `unexpected consequences' for the linearized observables. In particular, as we shall see below, the covariant Hamiltonian for the Jacobi system will explicitly depend on time.

An advantage of the new variable $n^a$ is that it in one-to-one correspondence with the position $x$ of the original system (at least, in the normal neighborhood of $\bar{x}$) while at the same time it is linear and belongs to a vector space. Therefore, one can perform an expansion in small $n^a$ and solve the system pertubatively.

The deviation vector $n^a$, \eqref{eq:nxtransf}, can be expressed as a derivative of the Synge world function\footnote{%
In this appendix we use ${\sigma}$ only for the Synge world function, not for the deviation parameter as we did in the main text.}
${\sigma(x|y)}$, \cite{Synge:book}. The world function is defined as a half of the square of geodesic distance, with the sign given by the causal character of the geodesic,
\begin{equation}\label{eq:sigmadef}
    \sigma(x|y) = \pm\frac12 (\text{geodesic distance between ${x}$ and ${y}$})^2\;.
\end{equation}
The deviation vector can be written as
\begin{equation}\label{eq:ncovdsigma}
    n^a = n^a(\bar{x}|x)\equiv-\bar{\nabla}^a\sigma(\bar{x}|x)\;,
\end{equation}
see, e.g., \cite{Synge:book,DeWittBrehme:1960,Christensen:1978yd}. Here we use the convention that ${\bar{\nabla}\sigma(\bar{x}|x)}$ denotes the derivative in the first argument ${\bar{x}}$ and ${\nabla\sigma(\bar{x}|x)}$ denotes the derivative in the second argument ${x}$. The normalization of $n^a$ is encoded in the relation
\begin{equation}\label{eq:nnorm}
    \sigma(\bar{x}|x) = \frac12\, n^a(\bar{x}|x)\,n^b(\bar{x}|x)\,\bar{g}_{ab} \;.
\end{equation}

When dealing with the Lagrangian, we also need a relation between velocities associated with variables ${x}$ and ${n}$. For that we assume that all variables in \eqref{eq:ncovdsigma} are time dependent, ${x(\tau)}$, ${\bar{x}(\tau)}$, and ${n^a(\tau)}$. Taking the covariant time derivative gives
\begin{equation}\label{eq:vurel}
    v^a\equiv\frac{D n^a}{D\tau} = -\bar{\nabla}^a\nabla_{\!b}\sigma(\bar{x}|x)\,u^b
       - \bar{u}^b\bar{\nabla}_{\!b}\bar{\nabla}^a\sigma(\bar{x}|x)\;.
\end{equation}
Let us notice, that in the coincidence limit $x=\bar{x}$, $u=\bar{u}$, i.e., setting $x$ and $u$ to the values of the central trajectory, we have $n=0$ and $v=0$ (cf.\ \eqref{eq:sigmacoinc} below).

The covariant expansion of the Lagrangian assumes that ${L}$ is written as a series in ${n}$ and ${v}$, and we ignore all terms of higher than second order,
\begin{equation}\label{eq:Lagrcovexp}
\begin{split}
    &L(x,u) = L_{0,0} + L_{1,0\,a}\,n^a + L_{0,1\,a}\,v^a\\
      &\qquad + \frac12 L_{2,0\,ab}\,n^a n^b + L_{1,1\,ab}\,n^a v^b + \frac12 L_{0,2\,ab}\,v^a v^b + \dots\;,
\end{split}\raisetag{7ex}
\end{equation}
with coefficients ${L_{k,l}}$ to be determined. For that, we need to take derivatives  ${\frac{\nabla}{\pa x}}$ and ${\frac{\pa}{\pa u}}$ of this relation, followed by the coincidence limit ${x=\bar{x}}$, ${u=\bar{u}}$.

First, we calculate the coincidence limit ${x=\bar{x}}$, ${u=\bar{u}}$ of derivatives of relations for $n(x)$ and $v(x,u)$  given by \eqref{eq:ncovdsigma} and \eqref{eq:vurel}. In this process we employ the following relations \cite{Synge:book,DeWittBrehme:1960,Christensen:1978yd}:
\begin{gather}
    \sigma|_{x=\bar{x}}=0\;,\notag\\
    \bar{\nabla}_{\!a}\sigma|_{x=\bar{x}}= \nabla_{\!a}\sigma|_{x=\bar{x}}=0\;,\notag\\
    \bar{\nabla}_{\!a}\bar{\nabla}_{\!b} \sigma|_{x=\bar{x}}
    = - \bar{\nabla}_{\!a}\nabla_{\!b} \sigma|_{x=\bar{x}}
    = \bar{g}_{ab}\;,\label{eq:sigmacoinc}\\
    \bar{\nabla}_{\!a}\bar{\nabla}_{\!b}\nabla_{\!c} \sigma|_{x=\bar{x}}
    =\bar{\nabla}_{\!a}\nabla_{\!b}\nabla_{\!c} \sigma|_{x=\bar{x}}
    = 0\;,\notag\\
    - \bar{\nabla}_{\!a}\!\bar{\nabla}_{\!b}\!\nabla_{\!c}\!\nabla_{\!d} \sigma|_{x=\bar{x}}
    = \bar{\nabla}_{\!a}\!\nabla_{\!b}\!\nabla_{\!c}\!\nabla_{\!d} \sigma|_{x=\bar{x}}
    = \frac13\bigl(\bar{R}_{acbd}+\bar{R}_{adbc}\bigr)\,.\notag
\end{gather}
This yields the following:
\begin{equation}\label{eq:dernv}
\begin{gathered}
    \frac{\nabla_{\!a}n^k}{\pa x}\bigg|_{x=\bar{x}} = \delta_a^k\;,\quad
    \frac{\pa v^k}{\pa u^{a}}\bigg|_{\substack{{x=\bar{x}}\\u=\bar{u}}} = \delta_a^k\;,\\
    \frac{\nabla_{\!a}}{\pa x}\frac{\nabla_{\!b}}{\pa x}u^k\bigg|_{\substack{{x=\bar{x}}\\u=\bar{u}}}
    = -\bar{u}^l \bar{R}_{l(a}{}^{k}{}_{b)} + \frac12\bar{u}^l\bar{R}_l{}^k{}_{ab}\;
\end{gathered}
\end{equation}
for the non-vanishing derivatives. Next, we notice that the only non-vanishing derivatives of Lagrangian \eqref{eq:geodLagr} to the second order are
\begin{equation}\label{eq:geodLagrder}
\begin{gathered}
    L(\bar{x},\bar{u}) = \bar{L} = -\frac12 m\;,\quad
    \frac{\pa L}{\pa u^a}(\bar{x},\bar{u}) = \bar{p}_a = m \bar{u}_a\;,\\
    \frac{\pa}{\pa u^a}\frac{\pa}{\pa u^b}L(\bar{x},\bar{u}) = m\bar{g}_{ab}\;.
\end{gathered}
\end{equation}
Employing \eqref{eq:dernv} and \eqref{eq:geodLagrder}, the coincidence limit of derivatives of the expansion \eqref{eq:Lagrcovexp} yields the coefficients $L_{k,l}$. The expansion then reads
\begin{equation}\label{eq:Lagrexp}
    L = \bar{L}
       + \bar{p}_a v^a
              +\frac{m}2\, \bar{g}_{ab}\, v^a v^b
       -\frac{m}2 \bar{u}^k \bar{u}^l \bar{R}_{kalb}\, n^a n^b + \dots\;.
\end{equation}

Let us take a closer look at the absolute and linear terms. We can do that for a general Lagrangian. The expansion \eqref{eq:Lagrexp} up to the first order in such a case gives
\begin{equation}
    L = L(\bar{x},\bar{u})
       + n^a \frac{\nabla_{\!a}L}{\pa x}(\bar{x},\bar{u})
       + v^a \frac{\pa L}{\pa u^a}(\bar{x},\bar{u})
       + \dots\;.
\end{equation}
If we use that the central trajectory satisfies the Euler--Lagrange equations
\begin{equation}\label{eq:eqmotbar}
    \frac{\nabla_{\!a}L}{\pa x}(\bar{x},\bar{u}) = \frac{D\bar{p}_a}{D \tau}\;,
\end{equation}
the definition of the canonical momentum $\bar{p}$,  and express the velocity as a derivative we get
\begin{equation}\label{eq:Lagrexplin}
    L-\bar{L} = \frac{D\bar{p}_a}{D \tau} n^a +  \bar{p}_a \frac{D n^a}{D \tau} = \frac{d}{d\tau}\bigl(\bar{p}_a n^a\bigr)\;.
\end{equation}
Thus, for an arbitrary Lagrangian the absolute and linear terms can be written as a total derivative term $\frac{d}{d\tau} f(n)$ with
\begin{equation}\label{eq:fndef}
    f(n) = \bar{S} + n^a \bar{p}_a \;,
\end{equation}
where ${\bar{S} = \int \bar{L} d\tau}$ is the action along the central trajectory.
This means that we can define the modified Lagrangian $\lagr$ as
\begin{equation}\label{eq:Lagrrel}
    L(x,u) = \frac{d}{d\tau} f(n) + \lagr(n,v) \;.
\end{equation}
Lagrangians which differ by a total derivative terms define the same dynamics. Therefore, $\lagr(n,v)$ is a suitable Lagrangian for the linearized system which, in the highest order is quadratic in $n$ and $v$.

In particular, for the geodesic Lagrangian the expansion \eqref{eq:Lagrexp} gives
\begin{equation}\label{eq:linLagr}
    \lagr(n,v) = \frac{m}2\, \bar{g}_{ab}\, v^a v^b
       -\frac{m}2 \bar{u}^k \bar{u}^l \bar{R}_{kalb}\, n^a n^b + \dots\;,
\end{equation}
which is the Lagrangian \eqref{eq:Lagrangian_vanHolten}.

\subsection{Canonical transformation and covariant expansion of the Hamiltonian}

In the main text, we have derived the covariant Hamiltonian $\ham$, \eqref{eq:covHam}, by the Legendre transformation of the covariant Lagrangian $\lagr(n,v)$, \eqref{eq:Lnv}. Let us now re-derive it {using directly the transformation between original variables $(x,\,p)$ of the full trajectory and linearized variables $(n,\,\pi)$. Since the transformation \eqref{eq:nxtransf} from $x$ to $n$ is time dependent, the corresponding canonical transformation is also time dependent. Moreover, the Lagrangian $\lagr$ for the linearized system differs from the original one by a total derivative term, \eqref{eq:Lagrrel}. For both these reasons the Hamiltonian} $\ham$ is not simply given by the second-order expansion of the full Hamiltonian $H$. Rather, one has to first find the new Hamiltonian, given by the canonical transformation, and only then expand this new Hamiltonian to the second order.

As well known in classical mechanics, the point transformation \eqref{eq:nxtransf} and the change of the Lagrangian \eqref{eq:Lagrrel} leads to the canonical transformation generated by the generating function
\begin{equation}\label{eq:genfcgen}
    G(x|\pi) = \pi_a n^a(\bar{x}|x) + f(n(\bar{x}|x))\;.
\end{equation}
Substituting \eqref{eq:fndef} yields 
\begin{equation}\label{eq:genfcgen2}
    G(x|\pi) = (\pi_a +\bar{p}_a)n^a(\bar{x}|x) + \bar{S}\;.
\end{equation}
Here, $\bar{p}_a$ is the original momentum \eqref{eq:geodmom} along the central geodesic and $\bar{S}=\int \bar{L} d\tau$ is the action evaluated along the central geodesic integrated up to time $\tau$. The canonical transformation is given by equations
\begin{equation}\label{eq:xpnpicantr}
\begin{aligned}
    p_a &= \frac{\pa G}{\pa x^a}(x|\pi)=  (\pi_k+\bar{p}_k)\,\nabla_{\!a} n^k(\bar{x}|x)\;,\\
    n^a &= \frac{\pa G}{\pa \pi_a}(x|\pi) = n^a(\bar{x}|x)\;.
\end{aligned}
\end{equation}
Since the generating function is explicitly $\tau$ dependent through $\bar{x}$, $\bar{p}$ and $\bar{S}$, the Hamiltonians of the original system and of the linearized system differ by $\frac{\pa G}{\pa\tau}$ term,
\begin{equation}\label{eq:Hhrel}
    \ham = H + \frac{\pa G}{\pa\tau}
    = H
    + \bar{u}^c \bar{\nabla}_{\!c} n^a(\bar{x}|x) (\pi_a+\bar{p}_a) + \bar{L}\;.
\end{equation}
Here, $\bar{\nabla}$ indicates the covariant derivative of $n^a(\bar{x}|x)$ in argument $\bar{x}$ and $\bar{L}=\frac{d\bar{S}}{d\tau}$ is the original Lagrangian evaluated on the central geodesic.\footnote{%
Trivial term $\bar{S}$ in generating function \eqref{eq:genfcgen2} and the related term $\bar{L}$ in \eqref{eq:Hhrel} cancels the zeroth order term in $H$ when expanding around the central geodesic. The central geodesic in the new variables corresponds to $n^a=0$, $\pi_a=0$ and thanks to substraction of the zeroth order term the new Hamiltonian vanishes, $\ham(0,0)=0$. Of course, $\bar{L}$ contributes to the explicit time dependence of $\ham$.}
We also used that $\frac{D\bar{p}}{D\tau}=0$ for the central geodesic.

Of course, $H$ and $\bar{\nabla}_{\!c}n^a$ have to be expanded up to the second order in variable $n^a$.
The relation \eqref{eq:xpnpicantr} between the canonical momentum $p_a$ associated with $x$ and the canonical momentum $\pi$ associated with $n$ reads
\begin{equation}\label{eq:ppirel}
    p_a = -\bigl(\bar{p}_k + \pi_k\bigr) \bar{\nabla}^k\nabla_{\!a}\sigma(\bar{x}|x)\,,
\end{equation}
or, inverting it,
\begin{equation}\label{eq:ppirel}
    \pi_k = - (\bar{\nabla}\nabla\sigma)^{\!-1}\!\!{}_{k}{}^a(\bar{x}|x)\, p_a - \bar{p}_k \,.
\end{equation}
Similar to \eqref{eq:dernv}, taking the coincidence limit ${x=\bar{x}}$, ${p=\bar{p}}$ of the derivatives of this relation, one obtains,
\begin{equation}\label{eq:derpi}
\begin{gathered}
    \frac{\pa \pi_k}{\pa p_{a}}\bigg|_{\substack{{x=\bar{x}}\\p=\bar{p}}} = \delta^a_k\;,\\
    \frac{\nabla_{\!a}}{\pa x}\frac{\nabla_{\!b}}{\pa x}\pi_k\bigg|_{\substack{{x=\bar{x}}\\p=\bar{p}}}
    = -\frac13\bar{p}_l \bar{R}_{k(a}{}^{l}{}_{b)} - \frac12\bar{p}_l\bar{R}_k{}^l{}_{ab}\;.
\end{gathered}
\end{equation}
Here, we have used relations
\begin{gather}
    (\bar{\nabla}\nabla\sigma)^{\!-1}\!\!{}_{k}{}^l\big|_{x=\bar{x}}=-\delta^l_k\;,\notag\\
    \nabla_{\!a}(\bar{\nabla}\nabla\sigma)^{\!-1}\!\!{}_{k}{}^l\big|_{x=\bar{x}}=0\;,\label{eq:sigmainvcoinc}\\
    \nabla_{\!a}\nabla_{\!b}
    (\bar{\nabla}\nabla\sigma)^{\!-1}\!\!{}_{k}{}^l\big|_{x=\bar{x}}
    = \frac13 \bar{R}_{k(a}{}^{l}{}_{b)} + \frac12\bar{R}_k{}^l{}_{ab}\;.\notag
\end{gather}
derivable from \eqref{eq:sigmacoinc}.

Expanding the geodesic Hamiltonian $H$, \eqref{eq:geodHam}, and the generating function $G$, \eqref{eq:ppirel}, in $n$ and $\pi$, a similar technique as in the previous section for the Lagrangian (employing \eqref{eq:derpi} in the process) gives
\begin{equation}\label{eq:Hamexp}
    H = \bar{H} +\pi_a \bar{u}^a
    +\frac1{2m} \bar{g}^{ab} \pi_a \pi_b + \frac{m}{6}\bar{u}^k\bar{u}^l \bar{R}_{kalb}\, n^a n^b + \dots\;,
\end{equation}
and
\begin{equation}\label{eq:Gexp}
    \frac{\pa G}{\pa\tau} = -\bar{H} -\pi_a \bar{u}^a
    + \frac{m}{3}\bar{u}^k\bar{u}^l \bar{R}_{kalb}\, n^a n^b + \dots\;,
\end{equation}
Finally, the Hamiltonian $\ham$, \eqref{eq:Hhrel}, for the linearized system  gives
\begin{equation}\label{eq:linHamexp}
    \ham = H + \frac{\pa G}{\pa\tau} =
    \frac1{2m} \bar{g}^{ab} \pi_a \pi_b + \frac{m}{2}\bar{u}^k\bar{u}^l \bar{R}_{kalb}\, n^a n^b + \dots\;,
\end{equation}
which is the Hamiltonian \eqref{eq:covHam}.

To summarize, due to the time dependence of the canonical transformation \eqref{eq:nxtransf}, {and due to time-dependent shift \eqref{eq:Lagrrel} of the Lagrangian, the new Hamiltonian $\ham$ for the linearized system is not conserved, despite the fact that the original Hamiltonian $H$ is a conserved quantity along the geodesic.} Clearly,
\begin{equation}\label{eq:hamder}
    \frac{d}{d\tau}\ham = \frac{\pa\ham}{\pa\tau}\;.
\end{equation}



%

\end{document}